\documentclass[aps,pre,twocolumn,amsmath,amssymb,amsfonts,floatfix,superscriptaddress,showkeys,showpacs]{revtex4-1}
\usepackage{dcolumn}  
\usepackage{multirow}
\usepackage[T1]{fontenc}
\usepackage[latin1]{inputenc}
\usepackage{dsfont}
\usepackage{verbatim}
\usepackage{bm} 
\usepackage{hyperref} 
\usepackage{cleveref} 
\usepackage{floatrow}
\usepackage{natbib}
\usepackage{graphicx}
\usepackage{epstopdf}
\usepackage{ifpdf} 
\usepackage{epsfig}
\usepackage{color}
\usepackage[usenames,dvipsnames]{xcolor}
\usepackage[caption=false]{subfig}
\usepackage[export]{adjustbox}
\usepackage[colorinlistoftodos]{todonotes}

\newcommand{\rmd}{{\mathrm{d}}}
\newcommand{\rme}{{\mathrm{e}}}
\newcommand{\icomplex}{\dot\iota}
\newcommand{\ellB}{\ell_{\mathrm{B}}}
\newcommand{\kBT}{k_{\mathrm{B}}T}

\begin{document}

\title{Rodlike counterions at heterogeneously charged surfaces}

\author{Ali \surname{Naji}}\thanks{Corresponding author -- Email: {a.naji@ipm.ir}}
\affiliation{School of Physics, Institute for Research in Fundamental Sciences (IPM), Tehran 19395-5531, Iran}

\author{Kasra \surname{Hejazi}}
\affiliation{Department of Physics, Sharif University of Technology,  P.O. Box 11155-9161, Tehran, Iran}
\affiliation{School of Physics, Institute for Research in Fundamental Sciences (IPM), Tehran 19395-5531, Iran}
\affiliation{Department of Physics, University of California, Santa Barbara, CA 93106, USA}

\author{Elnaz \surname{Mahgerefteh}}
\affiliation{School of Physics, Institute for Research in Fundamental Sciences (IPM), Tehran 19395-5531, Iran}

\author{Rudolf \surname{Podgornik}}
\affiliation{School of Physical Sciences and Kavli Institute for Theoretical Sciences, University of Chinese Academy of Sciences, Beijing 100049, China}
\affiliation{CAS Key Laboratory of Soft Matter Physics, Institute of Physics, Chinese Academy of Sciences (CAS), Beijing 100190, China}
\affiliation{Department of Physics, Faculty of Mathematics and Physics, University of Ljubljana, and
Department of Theoretical Physics, J. Stefan Institute, 1000 Ljubljana, Slovenia}

\begin{abstract}
We study the spatial and orientational distribution of rodlike counterions (such as mobile nanorods) as well as the effective interaction mediated by them between two plane-parallel surfaces that carry fixed (quenched) heterogeneous charge distributions. The rodlike counterions are assumed to have an internal charge distribution, specified by a multivalent monopolar moment and a finite quadrupolar moment, and the quenched surface charge is assumed to be randomly distributed with equal mean and variance on the two surfaces. While equally charged surfaces are known to repel within the traditional mean-field theories, the presence of multivalent counterions has been shown to cause attractive interactions between uniformly charged surfaces due to the prevalence of strong electrostatic couplings that grow rapidly with the counterion valency. We show that the combined effects due to  electrostatic correlations (caused by the coupling between the mean surface field and the multivalent, monopolar, charge valency of counterions) as well as the disorder-induced interactions (caused by the coupling between the surface disorder field and the quadrupolar moment of counterions) lead to much stronger attractive interactions between two randomly charged surfaces. The interaction profile turns out to be a nonmonotonic function of the intersurface separation, displaying an attractive minimum at relatively small separations, where the ensuing attraction can exceed the maximum strong-coupling attraction (produced by multivalent monopolar counterions between uniformly charged surfaces) by more than an order of magnitude. 
\end{abstract}


\maketitle

\section{Introduction}
\label{sec:intro}

Electrostatic interactions are ubiquitous in colloidal, soft and biological systems \cite{Israelachvili}, and have been the focus of intense study for decades, yielding ever more sophisticated frameworks for their detailed understanding. Throughout this long history, we have witnessed a consistent refinement and sophistication of the methodological approaches as well as fine-tuning of the physical models ever since the original formulation of the mean-field Poisson-Boltzmann (PB) theory and the primitive model of a Coulomb fluid (comprising mobile charged species in a base fluid) were introduced \cite{Edwards,holm,book,Wong,PhysToday,hoda_review,andelman-rev,perspective,Naji_PhysicaA,Naji2010}. In part, this is due to a constantly increasing resolution of the experimental setups that can detect minute variations in direct interactions between charged nano-/macromolecular surfaces in a Coulomb fluid \cite{RMP2010,confin,borkovec-rev} and, in part, it is due to the increased reliability of the simulation methods that have been able to incorporate ever more realistic features pertaining to experimentally accessible and observable systems \cite{Linse,espresso1,espresso2}. In fact, it was the advanced simulations  \cite{Linse1} that led to a paradigm shift, allowing for a simple conceptual framework to replace the mean-field picture for highly charged systems. This eventually led to advanced theories that could be applied to strongly coupled Coulomb systems \cite{Levin02,Shklovs02,RMP2010,book,holm,Podgornik89,Podgornik89b,Netz-orland,Netz01,AndrePRL,AndreEPJE,hoda_review,Naji_PhysicaA,Naji2010,perspective,asim,SCdressed1,SCdressed2,SCdressed3,NajiNetzEPJE,NajiArnold,Naji_CCT1,Naji_CCT2,Weeks,Weeks2,Jho1,arnoldholm,jho-prl,asim,Woon1,Woon2,Woon3,Woon4,JPCM2009,trizac,Santangelo,Hatlo-Lue,Forsman04,Burak04,Levin2009}. The old PB picture was then upgraded to a dichotomy between the weak- and the strong-coupling approaches, delimiting the exact behavior of a Coulomb system at any value of electrostatic coupling, relevant not only conceptually but also directly applicable to the interpretation of experiments \cite{Trefalt}. 

While these advances in upgrading the mean-field imagery of the PB theory for the primitive model of Coulomb fluids are interesting in themselves \cite{holm}, we are presently more concerned with generalizations of the basic physical models on which the PB formulation is based. The charged point-particle model for mobile ions in a Coulomb fluid neglects all ion-specific effects and includes only the ion valency, giving thus a {\em one-parameter model}, where the ions differ only in the amount of charge they bear. One straightforward way to amend this drawback, sharing some of the conceptual simplicity with the original PB theory, is to take into account the excess,  static, ionic dipolar polarizability of the ions \cite{Ben-Yaakov2011,Ben-Yaakov2011b,Frydel2011,Vincent,Holm-pol,Sahin2,Sahin3}. This can be further generalized by accounting for the next order quadrupolar polarizability  \cite{Slavchov1,Slavchov2}, which, interestingly enough, eliminates some of the pesky infinities appearing in the nonpolarizable case. Another train of thought is to scrutinize the effects of the dipolar moment in the ionic charge distribution \cite{andelman1,andelman2}, or even higher order multipoles such as the quadrupolar moment, relevant for extended ions such as charged nanorods \cite{Forsman06,Messina,bohinc1,bohinc-rev,bohinc2012,Bohinc2011rev,Woon1,Woon2,Woon3,Woon4,JPCM2009,perspective}.

Another important feature at the very basis of the PB theory is the assumption that the charge distributions on the nano-/macromolecular surfaces bounding a Coulomb fluid are known and controllable. This is often not the case as the surface charge distribution can heavily depend on the method of preparation of the sample, leading to a disordered component in the surface charge density \cite{science11}. The coupling between electrostatic interactions and charge disorder has been discussed in some important cases, including surfactant-coated surfaces \cite{Meyer,Meyer2,klein,klein1,klein2,kekicheff1,kekicheff2,kekicheff3}, random polyelectrolytes and polyampholytes \cite{kantor-disorder0,andelman-disorder}, neutral heteropolymers with sequence-specific polarizabilities \cite{Bing1,Bing2},  randomly charged and patchy planar boundaries \cite{Rabin,netz-disorder2,Lukatsky1,Lukatsky2,ali-rudi,rudiali,partial,Ghodrat1,Ghodrat2,Ghodrat3,Andelman2013,Andelman2016,Andelman2017,Dietrich2018}, nanoporous materials \cite{Hribar}, graphene layers \cite{Shklovski-07}, and also in the context of Casimir interactions predominantly relevant  at the nanoscale  \cite{disorder-PRL,jcp2010,pre2011,epje2012,jcp2012,speake,kim2,kim3}. In all these cases, surface charge distributions often show a disordered component of the {\em quenched} type (i.e., they are not thermalized with the surrounding Coulomb fluid), leaving its fingerprint on various system properties, in particular on effective intersurface/intermolecular forces.  

Electrostatic quadrupoles are important for charged rodlike molecules that are found in a relatively broad range of scales in the biological context: Short fragments of DNA \cite{Pelta,Bloom2,Raspaud,Li2015,Pfohl2001,Pfohl2003,Bohinc2011rev}, F-actin  \cite{Wong-F,Pfohl2003},  cellular scaffold microtubules \cite{Needleman,Baker,Pfohl2001}, and various types of viruses such as $fd$ and $M13$ filamentous phages \cite{Tang1996,Butler2003,Grelet} can be understood based on anionic rod models. Some of DNA condensing counterions such as the naturally occurring spermidine and spermine have been modeled as rodlike or linear flexible polycations  \cite{bohinc-rev,bohinc2012,Bohinc2011rev,Gimsa,Feuerstein1986,Feuerstein1990}. Other examples include synthetic stiff-chain polyelectrolytes \cite{Guilleaume2001,Patel2004}, charged dendritic nanorods \cite{Gossel2002,Gou2009}, liquid-crystalline rigid-core polyions \cite{Goossens2016,Cheng2010,Wu2013}, and charged coated gold nanorods \cite{Umar2013}.  Strong electrostatic interactions were actually first invoked for the rodlike tobacco mosaic virus already in the seminal work of Bernal and Fankuchen \cite{Bernal}, which is also one of the first cases of the application of the PB theory to biological systems  \cite{Review1}. 
Charged rodlike viruses and viruslike nanoparticles have been used to assemble functional materials \cite{assembly1,assembly2,assembly3,assembly4}. In particular, surface alignment of filamentous plant-virus-derived nanoparticles (PVN) have been accomplished to enable the formation of PVN films, whose formation is controlled by the strength of the PVN surface interaction \cite{assembly4}. A strong electrostatic attraction between the deposition surface and filamentous viruses enhances a stable film formation, as is clear from the effect of the ionic strength and the $p$H of the solution \cite{assembly3}, while monolayer assembly of $M13$ filamentous phages via competitive electrostatic interactions on a  polyelectrolyte-multilayer deposition surface was used to organize phage-based electrodes \cite{assembly1}. Because of the preparation method, the deposition surface for the assembly of functional materials exhibits a disordered field that often remains unaltered after the assembly or fabrication of the materials, thus remaining effectively quenched; a feature that needs to be explored in order to understand the details of the deposition process.  


Motivated by the numerous occurrences of charged rodlike particles in soft-/biomatter context,   
we were invariably led to investigate and clarify the electrostatic interactions engendered by a Coulomb fluid consisting of mobile rodlike counterions between charged bounding surfaces.  This particular angle was discussed in both weak- and strong-coupling regimes in a previous publication  \cite{JPCM2009} by incorporating multipolar effects due to the extended internal charge of rodlike counterions and assuming that the bounding surfaces are planar and carry uniform surface charge densities. In what follows, we will advance further and combine the analysis of rodlike charge multipoles next to heterogeneously charged, planar bounding surfaces, characterized not only by their mean  charge densities, but also a quenched disorder component in their charge distribution. 

The organization of the paper is as follows: Our model and theoretical framework are introduced in Section \ref{sec:framework}. The analytical aspects of our results are discussed in Section \ref{sec:plane-parallel} with further numerical analysis presented in Section \ref{sec:plane-parallel_num}. The paper is concluded in Section \ref{sec:conclusion}. 

 \begin{figure}[t!]
 \centering
\includegraphics[scale=0.11]{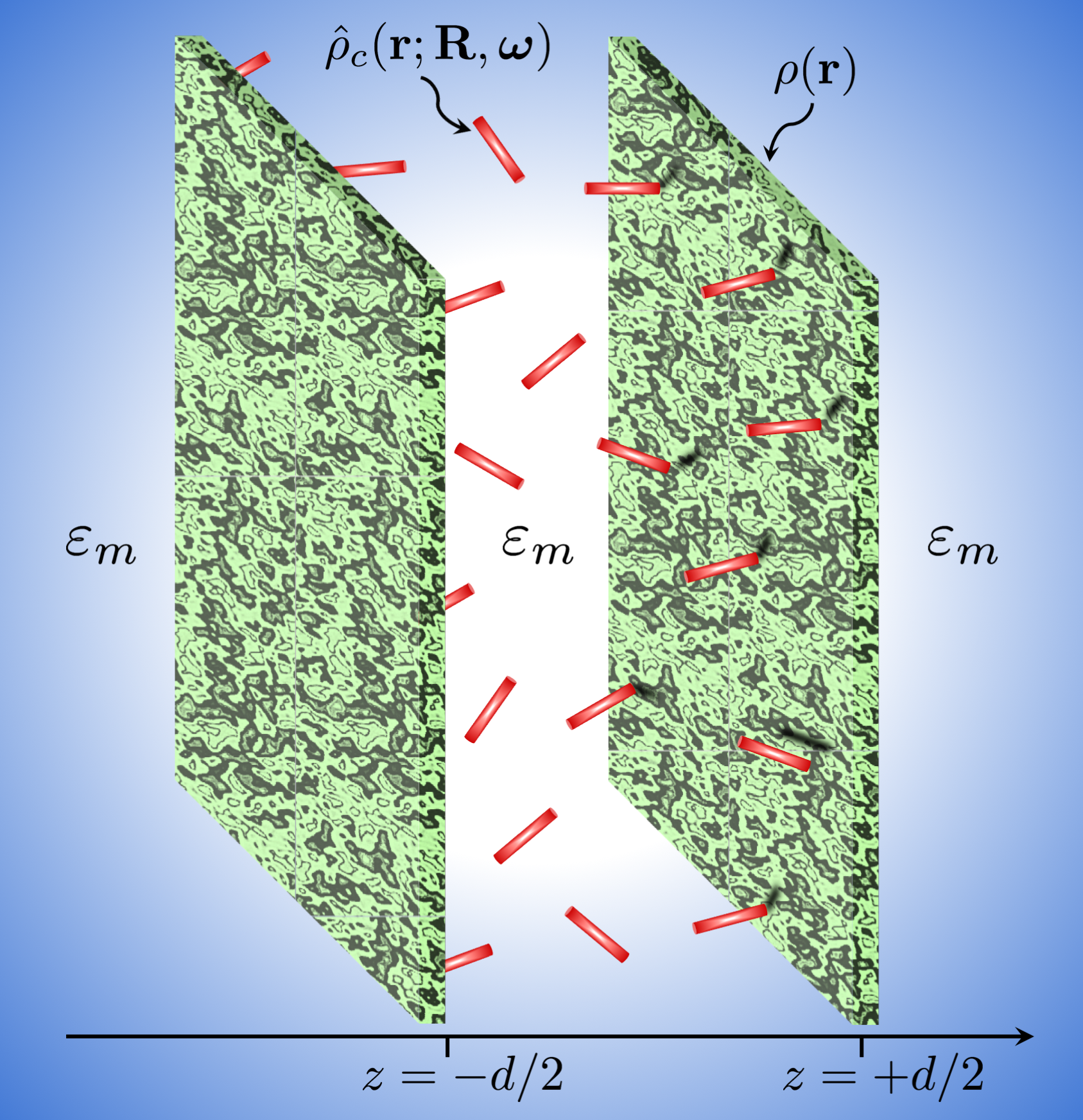}
\caption{Schematic view of mobile rodlike counterions between heterogeneously charged surfaces with quenched  random charge distribution $\rho(\mathbf{r})$. The counterions are typically multivalent, having an internal charge distribution $\hat{\rho}_{c}(\mathbf{r}; \mathbf{R}, \boldsymbol{\omega})$ with given monopolar and quadrupolar moments. 
}
\label{fig:schematic}
\end{figure}

\section{Model and Framework}
\label{sec:framework}

Our model consists of two fixed, charged, plane-parallel surfaces positioned normal to the $z$-axis at $z=\pm d/2$, with $d$ being the intersurface separation filled with a polarizable solvent (e.g., water), containing  charged mobile nanorods as rodlike surface `counterions' (see Fig. \ref{fig:schematic}). As our main goal is to elucidate the interplay between the multipolar nature of  the internal counterion-charge distribution and the surface charge heterogeneity, we make several assumptions to circumvent some of the inherent complexities of the problem, and also to facilitate an analytical approach, illuminating the underlying physics. 

First, we assume that the system is salt-free and that the dielectric constant $\varepsilon_m$ is uniform across the system, presenting no dielectric discontinuities across the bounding surfaces. Although this assumption presents a significant simplification (valid in situations where the charged surfaces have a small thickness \cite{jho-prl}), it nevertheless helps to disentangle the different competing factors involved by eliminating the more complex effects due to dielectric and salt ``image'' charges. The electrostatic interactions in the system in three dimensions can then be calculated using the bare, free-space, Coulomb kernel  
\begin{equation}
G_0(\mathbf{r},\mathbf{r}') = \frac{1}{4\pi\varepsilon_0\varepsilon_m |\mathbf{r}-\mathbf{r}'|}. 
\end{equation}
The image charge effects have extensively been discussed in other closely related cases by some of the present authors; see, e.g., Refs. \cite{SCdressed2,SCdressed3,perspective,JPCM2009,disorder-PRL,jcp2010,pre2011,epje2012,jcp2012,jho-prl,rudiali,Ghodrat1,Ghodrat2,Ghodrat3}. Our current formulation can straightforwardly be extended to include the dielectric/salt image effects; however, for the sake of clarity in our presentation, such an extension will be discussed elsewhere \cite{to-be-published1}. 
 
Secondly, the rodlike counterions are assumed to be structurally rigid with uniaxial, mirror symmetric, internal charge distribution with respect to their center of charge (of position vector $\mathbf{R}$), excluding thus any dipolar or odd-ordered multipolar moments. The tensorial quadrupolar moment is defined as $\boldsymbol{\hat Q}=t\,\mathbf{n}\otimes\mathbf{n}$ with  $\mathbf{n}$ being the particle director, parameterized by a set of angular variables $\boldsymbol{\omega}$. We further assume that the rodlike counterions are sufficiently short (as will be quantified later in Section \ref{subsec:validity_regime}), enabling a perturbative multipolar expansion for their single-particle, charge density operator at the arbitrary observation point $\mathbf{r}$ in space as
\begin{equation}
\hat{\rho}_{c}(\mathbf{r}; \mathbf{R}, \boldsymbol{\omega}) = e_0 \left[q\delta(\mathbf{r} - \mathbf{R}) +  t (\mathbf{n} \cdot \nabla)^2 \delta(\mathbf{r} - \mathbf{R}) + \cdots\right], 
\label{eq:one-particle-op}
\end{equation}
where $e_0$ is the elementary charge and $q$ is the counterion (monopolar) charge valency. Without any loss of generality, we  take $q> 0$. In Eq. \eqref{eq:one-particle-op}, we shall only deal with terms up to the quadrupolar order, as the higher-order terms (hexadecapole and above) turn out to be negligible within the regime of interest here (see Section \ref{subsec:validity_regime}).

Thirdly, we assume that the charge distribution on the bounding surfaces, $\rho(\mathbf{r})$, is {\em quenched} and spatially heterogeneous, and is given by the Gaussian probability weight, 
\begin{equation}
{\mathcal P}[\rho] = C_{\rho}\, \rme^{- \frac{1}{2}  \int {\mathrm{d}} {\mathbf r}{\mathrm{d}} {\mathbf r}'\,  [\rho({\mathbf r}) - \rho_0({\mathbf r})]{\mathcal G}^{-1}(\mathbf{r},\mathbf{r}')\,[\rho({\mathbf r}') - \rho_0({\mathbf r}')]}, 
\end{equation}
where $C_{\rho}$ is a normalization factor. We further assume that the disordered charge distributions on the two surfaces are statistically identical and independent, characterized  by equal mean surface charge densities, $\sigma$ (in units of $e_0$), and equal charge disorder variances,  $g\geq 0$  (in units of $e_0^2$), on the two surfaces. Hence, the mean and the two-point correlation functions 
\begin{align}
&\rho_0(\mathbf{r})\equiv \langle\!\langle\rho(\mathbf{r}) \rangle\!\rangle = - \sigma e_0 \left[ \delta\big(z-\frac{d}{2}\big) + \delta\big(z+\frac{d}{2}\big) \right], 
\label{eq:rho-mean}
\\
&{\mathcal G}(\mathbf{r},\mathbf{r}')\equiv \langle\!\langle\big( \rho(\mathbf{r}) - \rho_0(\mathbf{r}) \big)\big( \rho(\mathbf{r}') - \rho_0(\mathbf{r}') \big)\rangle\!\rangle 
\label{eq:rho-variance}\\
&\qquad=  ge_0^2 \left[ \delta\big(z-\frac{d}{2}\big) + \delta\big(z+\frac{d}{2}\big) \right]\!\delta(z - z'){\mathcal C}(\boldsymbol{\varrho} - \boldsymbol{\varrho}'). 
\nonumber
\end{align}
The brackets here represent ensemble average over different realizations of the quenched charge disorder, defined as $\langle\!\langle \cdots \rangle\!\rangle = \int {\mathcal{D}\rho}\,(\cdots)\,{\mathcal P}[\rho]$. The chosen form of the transverse disorder correlator, ${\mathcal C}(\boldsymbol{\varrho} - \boldsymbol{\varrho}')$, reflects the statistical homogeneity assumed in the transverse directions, $\boldsymbol{\varrho}=(x,y)$.  To model {\em correlated} surface disorder with finite in-plane correlation length $\xi$, we take a Yukawa-type transverse correlator as  
 \begin{equation}
 {\mathcal C}(\boldsymbol{\varrho} - \boldsymbol{\varrho}') = \frac{1}{2 \pi \xi^2}\,\mathrm{K}_0\!\left(\frac{|\boldsymbol{\varrho} - \boldsymbol{\varrho}'|}{\xi}\right),
\label{eq:Lorentzian}
 \end{equation}
where $\mathrm{K}_0(\cdot)$ is the zeroth-order modified Bessel function of the second kind. In Fourier space spanned by the transverse wavevector $\mathbf{k}$, we get the Lorentzian form  
\begin{equation}
 {\mathcal C}(k) = \frac{1}{\xi^2 k^2 + 1},\quad\quad k=|{\mathbf{k}}|. 
 \label{eq:Fourier_correlation_function}
\end{equation}
The case of {\em uncorrelated disorder}  is trivially reproduced by setting $\xi=0$, giving ${\mathcal C}(\boldsymbol{\varrho} - \boldsymbol{\varrho}')=\delta(\boldsymbol{\varrho} - \boldsymbol{\varrho}')$, or $ {\mathcal C}(k) =1$. 

Finally, since mobile nanorods are often multivalent, they are expected to cause significant non-mean-field effects. These can generally be quantified using a dimensionless electrostatic coupling parameter, which scales with the third power of the counterion valency (see Eq. \eqref{eq:Xi}) and, even for moderately large counterion valencies, its magnitude can be in few hundreds \cite{hoda_review,Naji_PhysicaA,Naji2010,perspective}. Despite its intriguing non-mean-field features, the behavior of Coulomb fluids at such large couplings has been shown to be described well (against exact analytical solutions \cite{exact1,exact2}, computer simulations \cite{book,holm,AndrePRL,AndreEPJE,hoda_review,Naji_PhysicaA,Naji2010,perspective,asim,SCdressed1,SCdressed2,SCdressed3,NajiNetzEPJE,NajiArnold,Naji_CCT1,Naji_CCT2,Weeks,Weeks2,Jho1,arnoldholm,jho-prl,asim,Woon1,Woon2,Woon3,Woon4}  and even experiments \cite{hoda_review,Trefalt}) using a virial expansion scheme, especially in its leading-order (single-particle) form, known as the  {\em strong-coupling theory}  \cite{Netz01}. This is the approach we will adopt here as well. 

Because of the quenched nature of the surface charge disorder, thermodynamic properties of the system can be calculated only after proper disorder-averaging is performed over the sample free energy of the system \cite{dotsenko1,dotsenko2}, which follows itself from the partition function obtained for a fixed realization of the quenched charge distribution, $\rho(\mathbf{r})$. This can be done most efficiently using a grand-canonical formulation, which, after the necessary analytical manipulations are done, can be transformed back to a canonical ensemble with $N$ counterions, ensuring the global electroneutrality condition  
\begin{equation}
2\sigma A = q N,  
\label{eq:ENC}
\end{equation}
 where $A$ is the (infinite) area of each bounding surface; note also that, because of our choice of $q>0$ and the sign convention in Eq. \eqref{eq:rho-mean}, we have  $\sigma> 0$. The aforementioned procedure is outlined in Appendix \ref{app:SC} with the disorder-averaged canonical free energy of the $N$-particle system obtained as 
\begin{align}
 {\mathcal F}_N&=  \frac{1}{2} \int \rmd\mathbf{r} \rmd\mathbf{r}'\,\rho_0(\mathbf{r})G_0(\mathbf{r},\mathbf{r}')\rho_0(\mathbf{r}')
\nonumber\\
&+ \frac{1}{2} \int\rmd\mathbf{r} \rmd\mathbf{r}'\, {\mathcal G}(\mathbf{r},\mathbf{r}')G_0(\mathbf{r},\mathbf{r}')
\nonumber\\
&- N \kBT\ln \int \rmd{\mathbf R} \rmd\boldsymbol{\omega}\, \Omega({\mathbf R}, \boldsymbol{\omega})\,  \rme^{ -\beta u\left({\mathbf R}, \boldsymbol{\omega} \right)},
\label{eq:F_N}
\end{align}
where $\beta=1/(\kBT)$, with $T$ being the absolute ambient temperature and $k_{\mathrm{B}}$ the Boltzmann constant. The angular integrals are taken over the unit sphere. The spatial integrals are taken over all space and the indicator function, $\Omega(\mathbf{r}, \boldsymbol{\omega})$, is defined such that it is one in the spatial regions accessible to counterions and zero otherwise. The first and second terms in Eq. \eqref{eq:F_N} give, respectively, the mean and the variance of the energy contributed by direct (bare) interactions between the surface charges (inclusive of the irrelevant self-interactions for each surface). The second term turns out to be a constant independent of the intersurface distance in the present context and, as such, will not be scrutinized any further (see Refs. \cite{rudiali,disorder-PRL,jcp2010,pre2011,epje2012,jcp2012} for the cases in which this term is of primary role). The last term in Eq. \eqref{eq:F_N} gives the free energy contribution of the multivalent counterions on the leading order. It involves the single-particle interaction energy between individual counterions and the surface charges, $u({\mathbf R}, \boldsymbol{\omega} ) $, which can be decomposed as 
\begin{equation}
u({\mathbf R}, \boldsymbol{\omega} ) = u^{(0)}({\mathbf R}, \boldsymbol{\omega} ) + u^{({\mathrm{dis}})}({\mathbf R}, \boldsymbol{\omega} ), 
\label{eq:u_general}
\end{equation}
with the first and second terms on the right-hand side giving the mean and variance of the single-particle energy due to the interactions between counterions and the surface charges \cite{Ghodrat1}. They are obtained as 
\begin{align}
&u^{(0)}\left({\mathbf R}, \boldsymbol{\omega} \right) = \int\rmd\mathbf{r} \rmd\mathbf{r}'\,\hat{\rho}_{c}(\mathbf{r}; {\mathbf R}, \boldsymbol{\omega})G_0(\mathbf{r},\mathbf{r}')\rho_0(\mathbf{r}'), 
\label{eq:u_0}
\\
&u^{({\mathrm{dis}})}\left({\mathbf R}, \boldsymbol{\omega} \right) = -\frac{\beta}{2}\int\rmd\mathbf{r} \rmd\mathbf{r}'\rmd\mathbf{r}'' \rmd\mathbf{r}'''\,\hat{\rho}_{c}(\mathbf{r}; {\mathbf R}, \boldsymbol{\omega})
G_0(\mathbf{r},\mathbf{r}')\nonumber\\
&\qquad\qquad\qquad\quad\times{\mathcal G}(\mathbf{r}',\mathbf{r}'')G_0(\mathbf{r}'',\mathbf{r}''')
\hat{\rho}_{c}(\mathbf{r}'''; {\mathbf R}, \boldsymbol{\omega}). 
\label{eq:u_dis}
\end{align}

We will also be   interested in the disorder-averaged number density profile of counterions,  
\begin{equation}
c({\mathbf r})  = \frac{N\int  \rmd\boldsymbol{\omega}\, \Omega({\mathbf r},\boldsymbol{\omega})\, \rme^{ -\beta u\left({\mathbf r}, \boldsymbol{\omega} \right)}}{\int \rmd{\mathbf R} \rmd\boldsymbol{\omega}\, \Omega({\mathbf R},\boldsymbol{\omega})\,  \rme^{ -\beta u\left({\mathbf R}, \boldsymbol{\omega} \right)}}, 
\label{eq:c_general}
\end{equation}
and in the orientational order of rodlike counterions that can be quantified using the tensorial quantity $3(\mathbf{n}\otimes\mathbf{n}-\mathbf{I})/2$ \cite{Lubensky}, where $\mathbf{I}$ is the unit matrix. 
This quantity needs to be thermally averaged and also averaged over different realizations of the quenched disorder, and then divided by $c({\mathbf r})$, to give the standardly defined orientational  order parameter $\mathbf{S}$. In the plane-parallel geometry considered here, we will be interested only in the nontrivial $zz$-component of $\mathbf{S}$, which,  in the polar representation with $n_z = \cos \theta$ and $\theta$ being the polar angle with respect to the $z$-axis, is expressed within the slit region as 
\begin{equation}
S_{zz}({\mathbf r})  = \frac{\int \! \rmd\boldsymbol{\omega} \left(\frac{3n_z^2}{2}-\frac{1}{2}\right) \rme^{ -\beta u\left({\mathbf r}, \boldsymbol{\omega} \right)}}{\int\! \rmd\boldsymbol{\omega}\,  \rme^{ -\beta u\left({\mathbf r}, \boldsymbol{\omega} \right)}}. 
\label{eq:S_general}
\end{equation}

Before proceeding with our discussion of the plane-parallel model, we note that the theoretical framework outlined above is applicable to any geometries for  fixed bounding surfaces carrying  quenched, Gaussian-distributed, disorder charges and is also applicable to any choice of rigid, internal, charge distribution for the mobile counterions  as long as the system is kept within the regime of validity of the current approach (Section \ref{subsec:validity_regime}).

\section{Analytical results}
\label{sec:plane-parallel}

\subsection{Dimensionless representation}
\label{subsec:dim_less}

The results to be discussed later for the plane-parallel model can be expressed in a dimensionless representation  obtained by dividing all energy scales with $\kBT$ and all lengthscales with the Gouy-Chapman length, 
\begin{equation}
\mu=\frac{1}{2\pi q \ellB\sigma}, 
\label{eq:mu}
\end{equation}
where $\ellB=e_0^2/(4\pi \varepsilon_m  \varepsilon_0 \kBT)$ is the Bjerrum length. The normal coordinate, the intersurface separation, and the in-plane disorder correlation length are  rescaled, respectively, as 
\begin{equation}
\tilde z  = \frac{z}{\mu},\,\,\,\,\, \tilde d  = \frac{d}{\mu},\,\,\,\,\, \tilde \xi  = \frac{\xi}{\mu}.
\end{equation}
We define the {\em effective half-length} of counterions as 
\begin{equation}
\ell\equiv \sqrt{\frac{2t}{q}}, 
\label{eq:ell_def_1}
\end{equation}
which is also rescaled as $\tilde \ell=\ell/\mu$, while the quadrupolar moment, $t$, is rescaled as 
\begin{equation}
\tilde t  = \frac{t}{q\mu^2} \equiv \frac{\tilde \ell^{\,2}}{2}. 
\label{eq:t_ell_def}
\end{equation}  
The effective length is in fact the only lengthscale associated with the internal structure of counterions that appears  in our model. For a charged rod, the effective length can generally be different from its {\em actual} end-to-end distance, as the former corresponds to the end-to-end length of an {\em equivalent dumbbell}, having the same mono-/quadrupolar moment as the charged rod (Appendix \ref{app:dumbbell}). 

The mean surface charge density, $\sigma$, and the surface charge variance, $g$, bring in other characteristic lengthscales, which lead to additional dimensionless parameters, i.e.,  the {\em (mean) electrostatic coupling parameter} \cite{Netz01}, 
\begin{equation}
\Xi= 2\pi q^3\ellB^{\,2}\, \sigma, 
\label{eq:Xi}
\end{equation}
which equals the rescaled Bjerrum length, $\Xi= q^2\ellB/\mu$, and the {\em disorder coupling (or strength) parameter} \cite{ali-rudi},
\begin{equation}
\chi = 2\pi q^2\ellB^{\,2}\, g. 
\label{eq:chi}
\end{equation}
The coupling parameter, $\Xi$, does not appear explicitly in our analytical expressions as the strong-coupling formulation used here formally corresponds to the leading-order theory obtained in the large $\Xi$ limit \cite{Netz01}; $\Xi$ will, however, appear explicitly, when the regime of applicability of the theory to realistic systems with finite values of $\Xi$ is considered (see Section \ref{subsec:validity_regime} and Appendix \ref{app:parameters}).  

Finally, the Fourier-transformed transverse disorder correlator in rescaled representation reads 
\begin{equation}
\tilde {\mathcal C}(\tilde k) = \frac{1}{\tilde \xi^2  \tilde k^2 + 1},  
\label{eq:C_k}
\end{equation}
while the counterion density profile, $c(z)$, the canonical free energy, ${\mathcal F}_N$, and the pressure $P$ exerted on the bounding surfaces are rescaled, respectively,  as 
\begin{equation}
\tilde c(\tilde z) = \frac{c(\mu \tilde z)}{2\pi \ellB \sigma^2},\,\,\,\, \tilde {\mathcal F}=\frac{\beta {\mathcal F}_N}{N },\,\,\,\,  \tilde P = \frac{\beta P}{2\pi \ellB \sigma^2}.
\label{eq:rescaled_defs}
\end{equation}

\subsection{General expressions}
\label{subsec:general_exp}

We now proceed by calculating the single-particle (counterion-surface) interaction energy, $u({\mathbf R}, \boldsymbol{\omega} ) $, defined through Eqs. \eqref{eq:u_general}-\eqref{eq:u_dis}. Focusing on the slit region between the two surfaces, where counterions are permitted to disperse, the single-particle interaction energy can be expressed  as 
\begin{equation}
\tilde{u}(\tilde z, \theta) = \tilde{u}^{(0)}(\tilde z, \theta)+ \tilde{u}^{({\mathrm{dis}})}(\tilde z, \theta).
\end{equation}
Using Eq. \eqref{eq:u_0}, $\tilde{u}^{(0)}$ is found to be a constant,  
\begin{equation}
\tilde{u}^{(0)}  = \tilde{d},  
\end{equation}
reflecting the fact that  the bare external electric field in the slit vanishes on average due to taking (equal) mean charge densities on the two surfaces. 

Using Eq. \eqref{eq:u_dis}, we find the contribution originating from the surface charge disorder, which can be decomposed into three different terms as
\begin{equation}
\tilde{u}^{({\mathrm{dis}})}(\tilde z, \theta) = \tilde{u}_{qq}(\tilde z)+2 \tilde{u}_{qt}(\tilde z, \theta)+\tilde{u}_{tt}(\tilde z, \theta).
\label{eq:three_u_terms}
\end{equation}
These terms can be calculated by expressing the spatial integrals over the transverse coordinate, $\boldsymbol{\varrho}$, as integrals over the transverse (Fourier) wavevector, ${\mathbf k}$, with rescaled norm $\tilde k=k\mu$. We thus find (i) the contribution due merely to the monopolar charge of the counterions, 
\begin{equation}
\tilde{u}_{qq}(\tilde z) = - \frac{\chi}{2} \int_0^{\infty} \frac{\rmd\tilde{k}}{ \tilde{k}} \, \tilde {\mathcal C}(\tilde{k}) \left[ \rme^{-2 \tilde{k} (\frac{\tilde{d}}{2} - \tilde{z})} +  \rme^{-2 \tilde{k} (\frac{\tilde{d}}{2} + \tilde{z})} \right],
\label{eq:u_qq_main}
\end{equation}
(ii)  the contribution originating from both monopolar and quadrupolar moments of the counterion charge,   
\begin{align}
\tilde{u}_{qt}(\tilde z, \theta) &= - \frac{\chi  \tilde{t}}{2}\left( \cos^2\theta - \frac{\sin^2\theta}{2} \right)
\label{eq:u_qt_main}
\\
&\quad\times \int_0^{\infty} \rmd\tilde{k} \, \tilde{k}  \, \tilde {\mathcal C}(\tilde{k}) \left[ \rme^{-2 \tilde{k} (\frac{\tilde{d}}{2} - \tilde{z})} +  \rme^{-2 \tilde{k} (\frac{\tilde{d}}{2} + \tilde{z})} \right],
\nonumber
\end{align}
and (iii) the contribution due purely to the quadrupolar moment of counterions,  
\begin{align}
\tilde{u}_{tt}(\tilde z, \theta)& = -  \frac{\chi   \tilde t^{\,2}}{2}\left( \cos^4\theta + \frac{1}{4}\sin^2\left(2\theta\right) + \frac{3}{8}\sin^4\theta \right)
\label{eq:u_tt_main}
\\
&\quad\times \int_0^{\infty} \rmd\tilde{k} \, \tilde{k}^3 \,  \tilde {\mathcal C}(\tilde{k}) \left[ \rme^{-2 \tilde{k} (\frac{\tilde{d}}{2} - \tilde{z})} +  \rme^{-2 \tilde{k} (\frac{\tilde{d}}{2} + \tilde{z})} \right]. 
\nonumber
\end{align}
It is evident that the contributions involving the quadrupolar moment (being the only terms that also depend on counterion orientation) will be present only if a finite degree of surface charge disorder is present. In other words, multipolar terms emerge only when the counterions experience nonuniform local fields. In any case, the disorder-induced terms \eqref{eq:u_qq_main}-\eqref{eq:u_tt_main} are found to be directly proportional to the disorder coupling strength. Now, defining the dimensionless integrals 
\begin{equation}
I_\nu(x) \equiv  \int_0^{\infty} \rmd\tilde{k}\,  \tilde{k}^{\nu-1} \tilde {\mathcal C}(\tilde{k}) \,\rme^{- \tilde{k}x},
\label{eq:I_nu}
\end{equation}
the above-mentioned terms can be expressed as
\begin{align}
&\tilde{u}_{qq}(\tilde z)  = - \frac{\chi}{2} \left[ I_0\left(\tilde{d} - 2 \tilde{z}\right) + I_0\left(\tilde{d} + 2 \tilde{z}\right) \right],
\label{eq:u_qq_exp}
\\
&\tilde{u}_{qt}(\tilde z, \theta) = - \frac{\chi\tilde t}{2} \left( \cos^2\theta - \frac{\sin^2\theta}{2} \right) 
\label{eq:u_qt_exp}
\\
&\qquad\qquad\qquad\quad\times\left[ I_2\left(\tilde{d} - 2 \tilde{z}\right) + I_2\left(\tilde{d} + 2 \tilde{z}\right) \right], 
\nonumber
\\
&\tilde{u}_{tt}(\tilde z, \theta) = - \frac{\chi\tilde t^{\,2}}{2}  \left( \cos^4\theta + \frac{1}{4}\sin^2\left(2\theta\right) + \frac{3}{8}\sin^4\theta \right) 
\label{eq:u_tt_exp}
\\
&\qquad\qquad\qquad\quad\times\left[ I_4\left(\tilde{d} - 2 \tilde{z}\right) + I_4\left(\tilde{d} + 2 \tilde{z}\right) \right]. 
\nonumber
\end{align}
Note that  $\tilde{u}_{qq}$, $\tilde{u}_{qt}$ and $\tilde{u}_{tt}$  depend explicitly on $\tilde{d}$, even though $\tilde{d}$ is not explicitly shown as an argument variable for them. 

The rescaled number density profile of counterions \eqref{eq:c_general} can now be evaluated as 
\begin{equation}
 \tilde{c}(\tilde{z}) =  \frac{2 \int_{-1}^{1} \rmd(\cos\theta)\,\tilde\Omega(\tilde{z},\theta)\, \rme^{- \tilde{u}(\tilde{z},\theta)}}{\int\! \rmd\tilde{z} \int_{-1}^{1}\rmd(\cos\theta)\,\tilde\Omega(\tilde{z},\theta)\, \rme^{- \tilde{u}(\tilde{z},\theta)}},
 \label{eq:c_z_final}
\end{equation}
where the indicator function $\tilde\Omega(\tilde{z},\theta)$ will be specified more precisely later (Section \ref{subsec:validity_regime}). The orientational order parameter \eqref{eq:S_general} is given by 
\begin{equation}
S(\tilde{z})  =  \frac{\int_{-1}^{1} \rmd(\cos\theta) P_2(\cos \theta)\,\rme^{- \tilde{u}(\tilde{z},\theta)}}{\int_{-1}^{1} \rmd(\cos\theta)\, \rme^{- \tilde{u}(\tilde{z},\theta)}},
\label{eq:S_plates}
\end{equation}
where $P_2(\cos \theta) = (3\cos^2 \theta-1)/2$ is the Legendre polynomial of second degree and we have omitted the $zz$ subscript from the previous notation used in Eq. \eqref{eq:S_general}. Finally,  the rescaled free energy of the plane-parallel system follows from Eq. \eqref{eq:F_N}  as 
\begin{equation}
\tilde {\mathcal F}=  - \frac{\tilde d}{2} - \ln \int\! \rmd\tilde{z} \, \rmd(\cos \theta)\, \tilde \Omega(\tilde z,\theta)\, \rme^{ - \tilde u\left(\tilde{z} , \theta \right)}, 
\label{eq:f_plates}
\end{equation}
giving the effective interaction pressure acting on the bounding surfaces (due to the disorder-induced and  the counterion-induced Coulomb interactions) as 
\begin{equation}
\tilde{P} = - 2 \frac{\partial \tilde{F}}{\partial\tilde{d}}.
\label{eq:p_plates}
\end{equation}

\subsection{Uncorrelated disorder: Surface singularities}
\label{subsec:uncorrelared_dis}

In the special case of uncorrelated surface charge disorder ($\tilde \xi=0$), we have $\tilde {\mathcal C}(\tilde k)=1$,  and     
\begin{equation}
I_\nu(x)  = x^{-\nu}\Gamma(\nu),
\end{equation}
where $\Gamma(\nu)$ is the Gamma function. In this case, the disorder-induced contributions to the single-particle interaction energy  follow in closed form by inserting the explicit form of $I_\nu(\cdot) $ in expressions \eqref{eq:u_qq_exp}-\eqref{eq:u_tt_exp}. Interestingly enough, these terms all turn out to be {\em singular} at the bounding surfaces located at  $\tilde z = \pm d/2$; that is, in an interval of sufficiently small size, $\tilde \zeta_\pm \equiv (\tilde{d}/2) \mp  \tilde{z}$ (note that $|\tilde z|<\tilde d/2$ and, hence, $\tilde \zeta_\pm>0$), we have 
\begin{align}
\tilde{u}_{qq}  \sim \frac{\chi}{2} \ln \tilde \zeta_\pm,\,\,\, \tilde{u}_{qt} \sim - \frac{\chi\tilde t}{2} \tilde \zeta_\pm^{-2},\,\,\,
\tilde{u}_{tt} \sim - \frac{\chi\tilde t^{\,2}}{2}\tilde \zeta_\pm^{-4}. 
\label{eq:singularities} 
\end{align}
These expressions also indicate that the counterion-surface interactions arising from the charge disorder are {\em attractive} and become shorter-ranged (and even more singular at the surfaces) at the higher orders of multipoles. Thus, rodlike counterions are expected to accumulate more strongly at randomly charged surfaces as compared with pointlike counterions of identical monopolar charge. 

The monopolar disorder-induced singularity leads to an {\em algebraic} divergence in the counterion density profiles,  $\tilde c(\tilde z)\sim \tilde \zeta_\pm^{-\chi/2}$, on approach to the boundaries, as discussed in detail elsewhere \cite{ali-rudi,Ghodrat1}. The multipolar terms $\tilde{u}_{qt}$ and $\tilde{u}_{tt}$, on the other hand, produce {\em essential} singularities of the forms $\sim\!\exp(\tilde \zeta_\pm^{-2})$ and $\sim\!\exp(\tilde \zeta_\pm^{-4})$, respectively,  in the counterion density at the boundaries. 

One must however note that, in our treatment of higher-order multipoles introduced initially through Eq. \eqref{eq:one-particle-op}, we have assumed that the higher-order terms are perturbatively small. Hence, we must ensure that the terms of growing multipolar order in the single-particle interaction (Eq. \eqref{eq:three_u_terms}) also become increasingly small. Equation \eqref{eq:singularities} indicates that this  condition is met for $\tilde t\,\tilde \zeta_\pm^{-2}<1$ or, using Eq. \eqref{eq:t_ell_def}, for
\begin{equation}
\tilde \zeta_\pm > \frac{\tilde \ell}{\sqrt{2}}. 
\label{eq:zeta_t}
\end{equation}

It is worth emphasizing that the singularities noted above originate directly from the charge disorder and disappear altogether if the disorder variance is set to zero. Given a finite degree of quenched disorder, the monopolar term (corresponding to the case of structureless pointlike counterions) and the higher-order multipolar terms lead to different (growing) orders of singularity, with the latter being systematically derived from the former.  On a formal level,  such singularities stem from the  quenched disorder average over the Coulomb potentials that are created in space by the randomly distributed surface charges; hence, $u^{({\mathrm{dis}})}$ can be shown to coincide with the sample-to-sample variance of the surface electrostatic potential (see Ref. \cite{Ghodrat1} for details), which thus brings in a higher-order Coulomb kernel (see Eq. \eqref{eq:u_dis}) and a respectively more short-ranged, effective counterion-surface interaction (thus, e.g., on approach to a randomly charged surface, the linear dependence of the Coulomb potential on distance, as anticipated for a uniformly charged surface, is replaced by a logarithmic one \cite{Ghodrat1}).

\subsection{Correlated disorder}
\label{subsec:correlared_dis}

For correlated surface charge disorder, we use the Lorentzian correlator  \eqref{eq:C_k}. In this case, the integrals $I_\nu(x)$ in Eq. \eqref{eq:I_nu} are given in terms of hypergeometric functions. We only require $I_0$, $I_2$ and $I_4$ that can be expressed using sine/cosine integrals, $\mathrm{Si}(x) = \int_0^{x} \rmd s\,\sin s/s$ and $\mathrm{Ci} (x) = -\int_x^{\infty} \rmd s\, \cos s/s$, as 
\begin{align}
&I_0(x) = - \ln \left( \frac{x}{\tilde \xi} \right) + \mathrm{Ci} \left(\frac{x}{ \tilde \xi} \right) \cos\left( \frac{x}{\tilde \xi} \right) 
\label{eq:I_0_xi}
\\
&\,\,\,\,\quad\qquad\qquad\qquad- \sin\left(\frac{x}{\tilde \xi} \right)\left( \frac{\pi}{2} -  \mathrm{Si}\left( \frac{x}{\tilde \xi} \right) \right), \nonumber\\
&I_2(x) = - \tilde\xi^{-2} \left(I_0(x)  +\ln \left( \frac{x}{\tilde \xi} \right)\right),\\
&I_4(x) 
= - \tilde\xi^{-4} \left(I_2(x)-\frac{\tilde \xi^2}{x^2} \right), 
\end{align}
with an irrelevant additive constant omitted in Eq. (\ref{eq:I_0_xi}). One can verify that $I_0(x)$ remains finite in the limit $x \to 0$, meaning that the {\em logarithmic} singularity due to the purely monopolar contribution, $\tilde{u}_{qq}$, is automatically  regularized, when the surface charge disorder has a finite in-plane correlation. The mixed mono-quadrupolar and the purely quandrupolar terms, $\tilde{u}_{qt}$ and $\tilde{u}_{tt}$, involving $I_2$ and $I_4$, respectively, still show singularities, albeit weaker ones as compared with their counterparts in the case of uncorrelated disorder, Eq. \eqref{eq:singularities}. In general, as $\tilde \xi$ is increased, the disorder effects decrease, and eventually vanish for $\xi\rightarrow \infty$, even if the disorder strength parameter, $\chi$, is fixed. 

\subsection{Regime of validity and choice of parameters}
\label{subsec:validity_regime}

Before proceeding further with our numerical analysis, which will require specific numerical values to be assigned to the system parameters, we shall first discuss the regime of applicability of the current approach over the parameter space. This can be achieved through the following validity criteria. 

{\em Strong-coupling criterion.--} The strong-coupling framework used here (see Section \ref{sec:framework} and  Appendix \ref{app:SC}) is based on a single-particle expression for the partition function of the system obtained on the leading order from a systematic virial and $1/\Xi$ expansion \cite{Netz01}. This mirrors the fact that, at large electrostatic coupling strengths, $\Xi\gg 1$, counterions are isolated in large correlation holes within  strongly correlated  (or, even, Wigner crystalline) quasi-two-dimensional layers they form over (oppositely) charged surface boundaries. Being confirmed by extensive numerical simulations \cite{RMP2010,book,holm,AndrePRL,AndreEPJE,hoda_review,Naji_PhysicaA,Naji2010,perspective,asim,SCdressed1,SCdressed2,SCdressed3,NajiNetzEPJE,NajiArnold,Naji_CCT1,Naji_CCT2,Weeks,Weeks2,Jho1,arnoldholm,jho-prl,asim,Woon1,Woon2,Woon3,Woon4}, this picture implies that the limiting theory for the two-surface geometry remains valid as long as the intersurface separation, $d$, is smaller that the lateral spacing, $2a_\perp$, between counterions. This lateral spacing is given approximately as $a_\perp \simeq \left(q/2\pi\sigma\right)^{1/2}$ or, in rescaled units, $\tilde a_\perp \simeq \sqrt{\Xi}$. This leads to the well-established criterion that the limiting theory can safely be used at finite electrostatic couplings, provided that $d<2a_\perp$, or
\begin{equation}
   \tilde d < 2\sqrt{\Xi}.  
   \label{eq:sc}
\end{equation}
Outside this regime of validity, the  finite-$\Xi$ corrections to the limiting theory become important \cite{AndrePRL,AndreEPJE,trizac}. 

\floatsetup[figure]{style=plain,subcapbesideposition=top}
\begin{figure*}[t!]
\centering
  \hspace{-2mm}
\sidesubfloat[]{%
  \label{fig:density_profile_a}
  \hspace{-7mm}
  \includegraphics[width=0.32\linewidth]{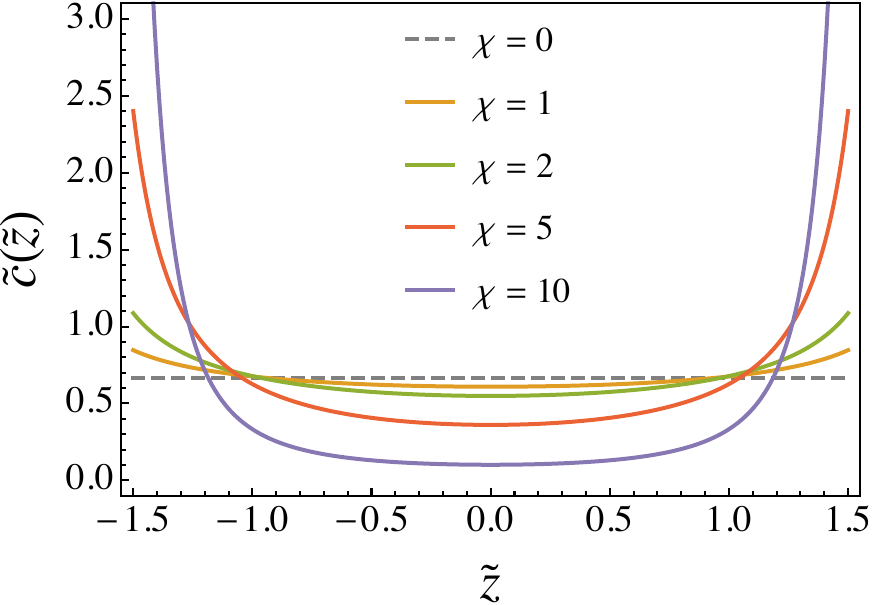}%
  \hspace{0mm}
}
\sidesubfloat[]{%
  \label{fig:density_profile_b}
    \hspace{-7mm}
\includegraphics[width=0.32\linewidth]{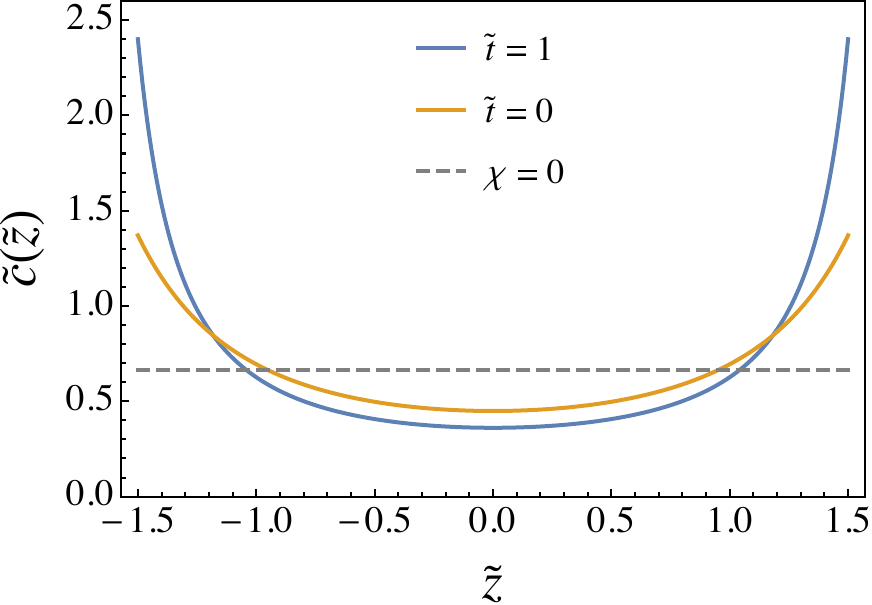}%
  \hspace{1mm}
  }
\sidesubfloat[]{%
  \label{fig:density_profile_c}
    \hspace{-8mm}
\includegraphics[width=0.32\linewidth]{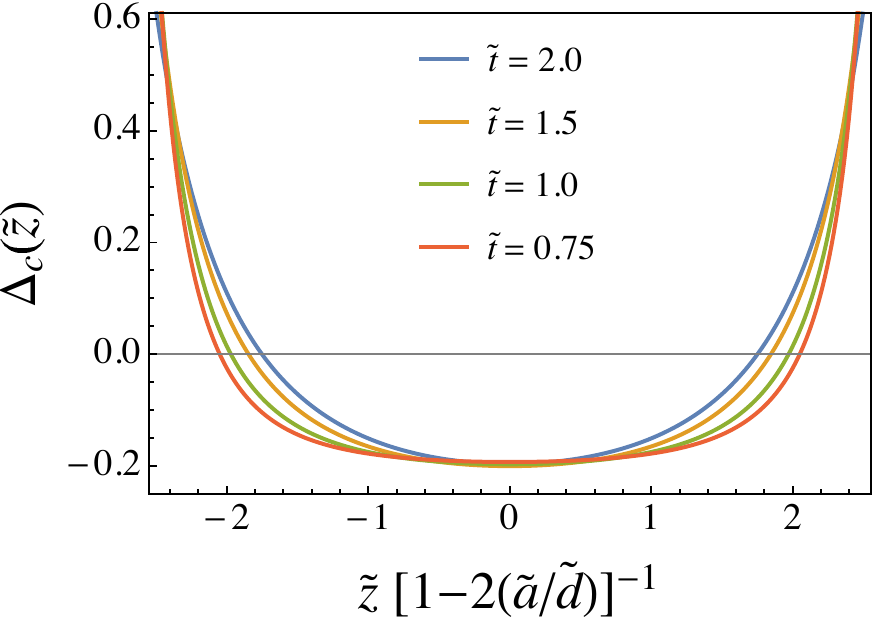}%
  }
\caption{(a) Rescaled density profile of rodlike counterions as a function of the rescaled normal coordinate $\tilde z$ (within the accessibility range $|\tilde z|<\tilde d/2-\tilde a$) between two randomly charged surfaces with no in-plane disorder correlation ($\tilde \xi=0$), and for fixed $\tilde d=5$, $\tilde t =1$ (hence, $\tilde a =1$; see Eq. \eqref{eq:cutoff}),  and  different values of $\chi$, as indicated on the graph. The reference case of counterions between uniformly charged surfaces ($\chi=0$, Eq. \eqref{eq:c0}) is shown by gray dashed line. (b) Same as panel (a) but here we compare the three cases $\tilde t=1$, $\tilde t=0$ (see Eq. \eqref{eq:c_t_0}), where $\chi=5$ is fixed, and $\chi=0$;  in all three cases, $\tilde d=5$ and $\tilde a =1$. (c) The relative density difference, $\Delta_c(\tilde z)$, see Eq. \eqref{eq:diff_c}, for fixed $\chi=5$, $\tilde d=5$,  and different values of $\tilde t$, as indicated on the graph, where $\tilde a$ is  suitably adjusted through Eq. \eqref{eq:cutoff}. 
}
\label{fig:density_profile}
\end{figure*}

{\em Multipole-expansion criterion.--} The condition \eqref{eq:zeta_t} for the validity of the multipole expansion scheme can be satisfied by adopting a finite {\em cutoff} or {\em closest-approach distance}, $a$, from the boundaries in such a way that 
\begin{equation}
\frac{\tilde \ell}{\sqrt{2}}\leq  \tilde a <\tilde \zeta_\pm.
\label{eq:ell_less_a} 
\end{equation}
This means that the rescaled indicator function, which is used to confine the counterions to the slit region between the surfaces, is to be defined as
\begin{align}
\tilde \Omega(\tilde z,\theta)\rightarrow \tilde \Omega\left(\tilde z,\theta; \frac{\tilde a}{\tilde d}\right) = \Theta\!\left(\frac{\tilde a}{\tilde d}-| \tilde z|\right), 
 \label{eq:indicator}
\end{align}
where $\Theta(\cdot)$ is the Heaviside step function. This ensures that counterions do not enter the interfacial regions $\tilde \zeta_\pm<\tilde a$ and the essential singularities noted above are properly regularized. The condition  \eqref{eq:ell_less_a} also justifies the assumption made through Eq. \eqref{eq:indicator} that the indicator function is independent of the counterion orientation. 

The two preceding criteria, Eqs. \eqref{eq:sc} and \eqref{eq:ell_less_a},  imply that the intersurface distance is bounded from below and above as $2a\leq d<2a_\perp$, or $2\tilde a\leq \tilde d<2\sqrt{\Xi}$.

{\em Gaussian criterion.--} In taking a Gaussian distribution for the quenched charge disorder on the bounding surfaces, we have implicitly assumed that the variations of the surface charge around its mean is small (see Ref. \cite{partial}). This assumption generally agrees with the multipole-expansion criterion as well, because the higher-order mutipoles are more strongly enhanced as the disorder strength is increased (Section \ref{subsec:general_exp}). This sets another validity criterion for the current approach as
\begin{equation}
\frac{g}{\sigma} = \frac{q\chi}{\Xi}<1.
\label{eq:gauss} 
\end{equation}

{\em Excluded-volume criterion.--} Due to their strong accumulation at the charged surfaces, multivalent (rodlike)  counterions, which in actual systems will be of finite length, could interact sterically, when their interspacing is smaller than their size. Thus, being also ignored in our model, such excluded-volume effects can be avoided by taking $\ell<a_\perp$ or, in rescaled units, 
\begin{equation}
 \tilde \ell <\sqrt{\Xi}.
\label{eq:steric} 
\end{equation}
This condition can easily be satisfied since the closest-approach distance will have to be chosen to fulfill Eq. \eqref{eq:ell_less_a} (see Appendix \ref{app:parameters}). 

According to the above criteria, the different lengthscales of the system are then ordered as $\sqrt{2}\ell\leq 2a\leq d<2a_\perp$, or in rescaled units, 
\begin{equation}
\sqrt{2}\tilde \ell\leq  2\tilde a\leq \tilde d<2\sqrt{\Xi}.
\label{eq:scales_ordered} 
\end{equation}

In what follows, the rescaled parameters are varied in such a way as to satisfy the above criteria and also to represent  real-life examples of rodlike counterions, when transformed back to actual units, as further discussed in Appendix \ref{app:parameters}. In particular, the disorder coupling parameter is increased from $\chi=0$ up to around (or slightly above) 10, the rescaled quadrupolar moment of rodlike counterions from $\tilde t=0$ up to 2 (or,  according to Eq. \eqref{eq:t_ell_def}, their effective half-length from $\tilde \ell=0$ up to 2), and the rescaled intersurface distance from its minimum value of $\tilde d=2\tilde a$ up to around $7\tilde a$. The rescaled in-plane correlation length, $\tilde \xi$, is varied over a wide range of values. The closest-approach distance will be set equal to its minimum admissible value (see Eq. \eqref{eq:ell_less_a}) as
\begin{equation}
\tilde a=\frac{\tilde \ell}{\sqrt{2}}=\sqrt{\tilde t}. 
\label{eq:cutoff}
\end{equation}

\section{Numerical results}
\label{sec:plane-parallel_num}

\subsection{Counterion density profile}

We first concentrate on the case of rodlike counterions confined between two planar surfaces covered with {\em uncorrelated} disordered charge distributions ($\tilde \xi=0$). The density profiles of counterions can be computed using Eq. \eqref{eq:c_z_final} and the appropriate expressions from Sections \ref{subsec:general_exp} and \ref{subsec:uncorrelared_dis}. The results are shown in dimensionless form in Fig. \ref{fig:density_profile}\subref{fig:density_profile_a} for fixed $\tilde d=5$, $\tilde t =1$ (hence, $\tilde a =1$ through Eq. \eqref{eq:cutoff}),  and  different values of $\chi = 0,\cdots,10$ (with the corresponding curves appearing from top to bottom at $\tilde z=0$). The density profiles are shown only within the interval $|\tilde z|<\tilde d/2 - \tilde a$, being accessible to counterions. 

The {\em reference case} of rodlike counterions between {\em uniformly} charged surfaces with $\chi=0$ is shown as gray dashed line. The (strong-coupling) density profile in this  case is spatially uniform across the slit region between the surfaces and is straightforwardly obtained as 
\begin{equation}
\tilde c_0(\tilde z)=\frac{2}{\tilde d-2\tilde a}\, \Theta\!\left(\frac{\tilde a}{\tilde d}-| \tilde z|\right). 
\label{eq:c0}
\end{equation}
The even distribution of counterions in the slit in this case corroborates the previously obtained result \cite{JPCM2009} that higher-order   (i.e., beyond monopolar)  moments of the internal  charge distribution of counterions make no contribution to the strong-coupling behavior of the system, unless there are field sources (e.g., dielectric image charges as in Ref. \cite{JPCM2009}, or surface charge disorder, as in the present context) that create spatially varying external potentials within the slit. 

As seen in the plot, counterions are accumulated near the charged surfaces, at the outer extremities of the shown interval at $|\tilde z_s|=\tilde d/2 - \tilde a=1.5$, and are depleted from the midplane at $\tilde z=0$, when the surfaces possess a nonvanishing degree of charge disorder $\chi>0$. The `contact' density $\tilde c(\pm \tilde z_s)$ is increased by 100\% for moderately large disorder coupling parameters $\chi\sim 2$, and rapidly grows as $\chi$ is further increased. This reflects the singular nature of the disorder-induced attractive interactions that enter through $\tilde{u}^{({\mathrm{dis}})}$, Eq. \eqref{eq:three_u_terms}, and diverge at the position of the two surfaces. 

To disentangle the effects due to the surface charge disorder and those from the quadrupolar moment of counterions, we show the density profile of counterions in Fig. \ref{fig:density_profile}\subref{fig:density_profile_b} for fixed $\chi=5$, $\tilde d=5$ and $\tilde a =1$, by considering three separate cases: $\tilde t=1$ (reproduced from panel \subref{fig:density_profile_a}), the special case with $\tilde t=0$ (where quadrupolar effects are set to zero, while the disorder effects are still included through the monopolar counterion-surface interaction terms), and the reference case with  $\chi=0$ (where both quadrupolar and disorder effects are absent). The deviations between the two latter cases (compare orange solid curve and gray dashed line) represent the effects due merely to the coupling between the monopolar moment of counterions and the disordered charges on the two surfaces, through the term $\tilde{u}_{qq}$ in Eq. \eqref{eq:three_u_terms}. The density profile for $\tilde t=0$ (corresponding, e.g., to spherical counterions of radius $a$) in the slit can be obtained as 
\begin{equation}
\tilde c(\tilde z)\bigg|_{\tilde t=0}=\frac{2C_0^{-1}(\chi,\tilde a/\tilde d)}{\big(\tilde d-2\tilde a\big)}\left(\frac{1}{4}-\frac{\tilde z^2}{\tilde d^{\,2}}\right)^{\!-\chi/2}\! \Theta\!\left(\frac{\tilde a}{\tilde d}-| \tilde z|\right), 
\label{eq:c_t_0}
\end{equation}
where $C_0(\chi,\tilde a/\tilde d)=2^{\chi} \,{}_2F_1\big(\frac{1}{2}, \frac{\chi}{2}, \frac{3}{2}, (1-\frac{2\tilde a}{\tilde d})^2\big)$, with special values $C_0(0,\tilde a/\tilde d)=1$, leading to Eq. (\ref{eq:c0}), and $C_0(\chi,0)=2^{-1+\chi}\sqrt{\pi}\,\Gamma\left(1-\chi/2\right)/\Gamma\left(3/2-\chi/2\right)$, reproducing the previously obtained result for pointlike counterions in Refs. \cite{ali-rudi,Ghodrat3}. 

Figure \ref{fig:density_profile}\subref{fig:density_profile_b} also shows that, although the surface accumulation (or, midplane depletion) of counterion is highly enhanced by the monopolar term alone, the combined effects due to the cooperation between the surface charge disorder and the quadrupolar moment of counterions, entering through the terms $\tilde{u}_{qt}$ and $\tilde{u}_{tt}$ in Eq. \eqref{eq:three_u_terms}, can lead to a sizably larger effect (compare blue and orange curves). This can be quantified using the quantity 
\begin{equation}
\Delta_c(\tilde z)\equiv \frac{c(\tilde z) - \tilde c(\tilde z)|_{\tilde t=0}}{\tilde c(\tilde z)|_{\tilde t=0}},
\label{eq:diff_c}
\end{equation} 
which is plotted in Figure \ref{fig:density_profile}\subref{fig:density_profile_c} for fixed $\chi=5$, $\tilde d=5$,  and for different values of $\tilde t$, where $\tilde a$ is  suitably adjusted through Eq. \eqref{eq:cutoff}. The regions of (midplane) depletion with $\Delta_c(\tilde z)<0$ and (near-surface) enhancement of counterion density  with $\Delta_c(\tilde z)>0$ are clearly discerned. It is interesting to note that different curves plotted for  $\Delta_c(\tilde z)$ merge both at the midplane region and close to the boundaries, and indicate that quandrupole-dependent effects arise mainly within the narrow intermediate regions, where  $\Delta_c(\tilde z)$ changes sign. 
 
\floatsetup[figure]{style=plain,subcapbesideposition=top}
\begin{figure}[t!]
\centering
\sidesubfloat[]{%
  \label{fig:orientation_profile_a}
  \hspace{-5mm}
  \includegraphics[width=0.75\linewidth]{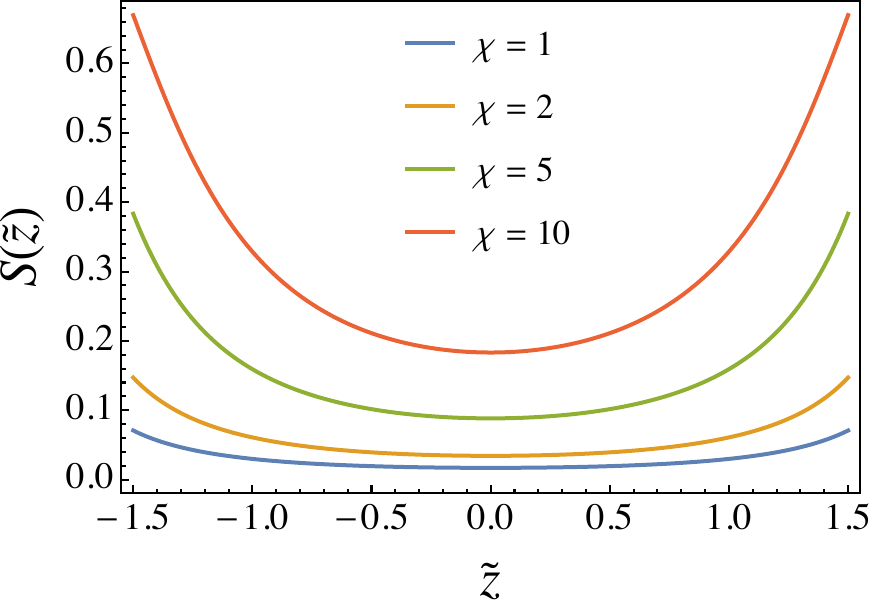}%
  }\\
  \vspace{3mm}
  \sidesubfloat[]{%
  \label{fig:orientation_profile_b}
  \hspace{-7mm}
  \includegraphics[width=0.75\linewidth]{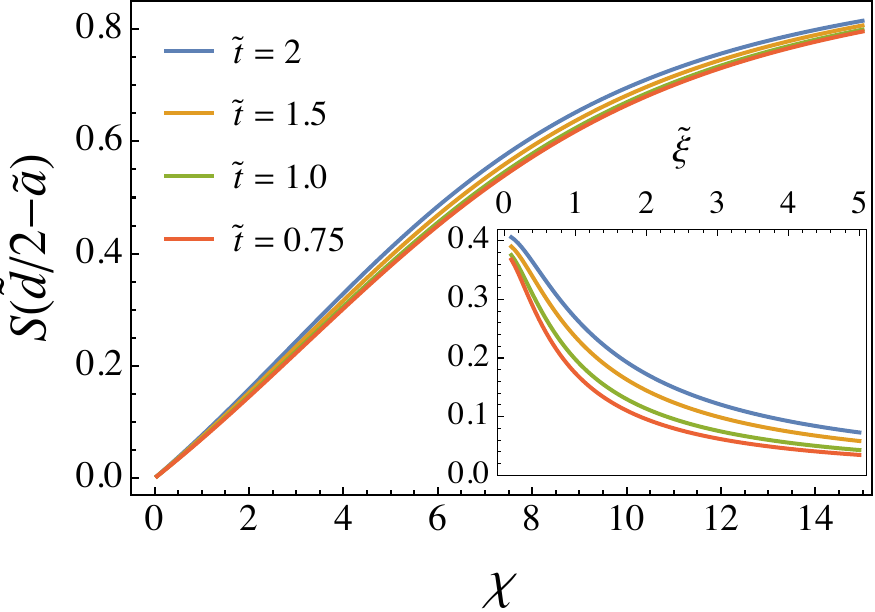}%
  }
\caption{(a) Orientational order parameter, $S$, of rodlike counterions as a function of the rescaled normal coordinate $\tilde z$ between two randomly charged surfaces with no in-plane disorder correlation ($\tilde \xi=0$), and for fixed $\tilde d=5$, $\tilde t =1$ ($\tilde a =1$),  and  different values of $\chi$, as indicated on the graph. (b) Same as (a)  but  $S$ is plotted at the closest-approach distance to the surfaces $|\tilde z_s|=\tilde d/2-\tilde a$ as a function of $\chi$ for different values of $\tilde t$ and fixed $\tilde d=5$, with suitably adjusted $\tilde a$, Eq. \eqref{eq:cutoff}. Inset shows the same quantity as a function of the rescaled  disorder correlation length, $\tilde \xi$, where we have fixed $\chi=5$.
}
\label{fig:orientation}
\end{figure}

\subsection{Counterion orientation profile}

The orientational order parameter $S(\tilde z)$, as computed from Eq. \eqref{eq:S_plates} and shown in Fig. \ref{fig:orientation}\subref{fig:orientation_profile_a}, also exhibits elevated values near the bounding surfaces as $\chi$ is increased, while it takes consistently smaller values around the midplane region. This quantity is generally bracketed by its limiting values as $-1/2\leq S\leq 1$, with $S=-1/2$ and 1 representing {\em nematic} orderings of the rodlike counterions {\em parallel} and {\em perpendicular} to the planar surfaces, respectively, and $S= 0$, representing an {\em isotropic} orientational distribution. In the reference case of  uniformly charged surfaces ($\chi=0$), one finds isotropic counterion orientation, $S= 0$, across the slit region (not shown), due again to the fact that, in this case, quadrupolar effects do not play a role in the strong-coupling regime \cite{JPCM2009}. The  orientation of rodlike counterions for disordered surfaces appears to remain nearly isotropic for small to moderately large values of $\chi$, especially in and around the midplane region. Closer to the bounding surfaces and/or for sufficiently large $\chi$, as shown in Figs. \ref{fig:orientation}\subref{fig:orientation_profile_a} and \ref{fig:orientation}\subref{fig:orientation_profile_b}, the order parameter takes larger positive values approaching one. This means that the combined effects due to the disorder-induced and the quadrupole-induced interactions tend to orient the rodlike counterions along the $z$-axis {\em perpendicular} to the planar surfaces. This physical picture is qualitatively different from the one obtained for rodlike counterions between uniformly charged bounding surfaces with  dielectric image charges, where counterions predominantly align  {\em parallel} to the surfaces \cite{JPCM2009}. 

When the charge disorder on the bounding surfaces has a finite correlation length $\tilde \xi$, our numerical results support the general observation in Section \ref{subsec:correlared_dis} that the disorder effects gradually diminish as the correlation length is increased to infinity. Consequently, the counterion density profile becomes increasingly more even (not shown) and the counterion orientation (Fig. \ref{fig:orientation}\subref{fig:orientation_profile_b}, inset) becomes more isotropic, as $S$ drops to smaller values,  tending to zero as $\tilde \xi$ is increased to infinity. 

\subsection{Effective interaction pressure}
\label{subsec:pressure}

\begin{figure}[t!]
\centering
\sidesubfloat[]{%
  \label{fig:pressure_a}
  \hspace{-6mm}
  \includegraphics[width=0.75\linewidth]{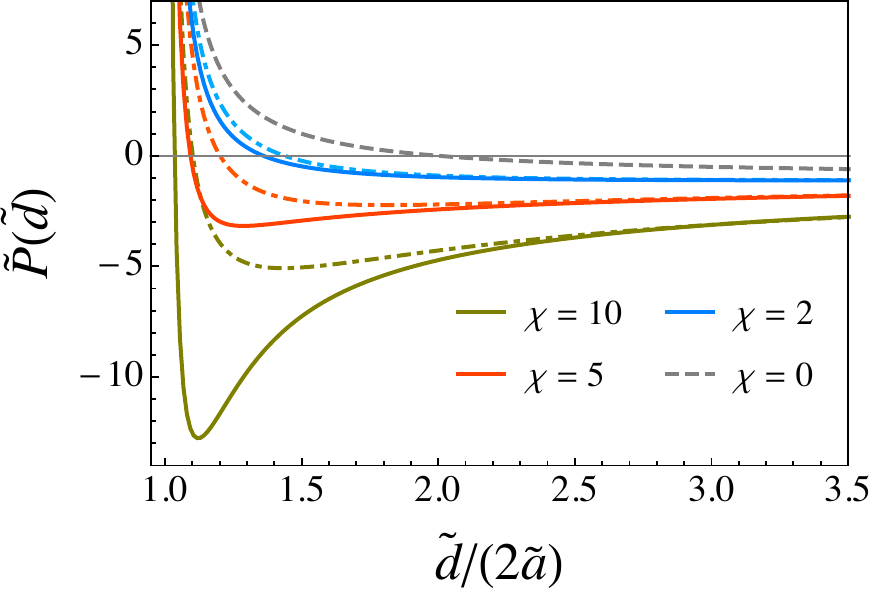}%
  }\\
  \vspace{3mm}
  \sidesubfloat[]{%
  \label{fig:pressure_b}
  \hspace{-3mm}
  \includegraphics[width=0.75\linewidth]{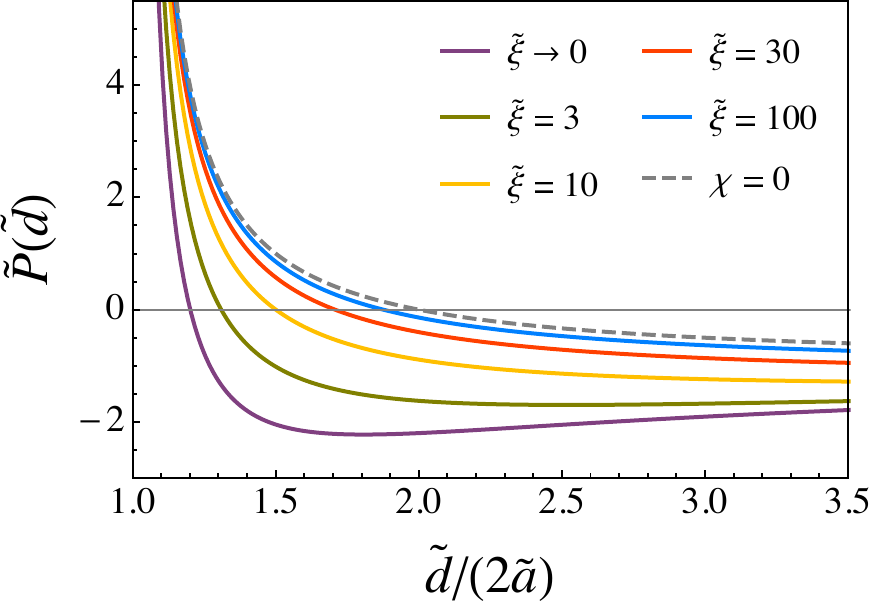}%
  }
\caption{(a) Rescaled effective pressure acting on the bounding surfaces carrying uncorrelated disorder charge ($\tilde \xi=0$) as a function of the rescaled intersurface distance $\tilde d/(2\tilde a)$ for fixed $\tilde t =1$ ($\tilde a =1$),  and  different values of $\chi$, as indicated on the graph (colored solid curves). The corresponding cases with $\tilde t=0$ are shown as colored dashed curves. (b) Same as (a) but for fixed $\chi=5$, $\tilde t =1$, and different values of the rescaled disorder correlation length, $\tilde \xi$.  In (a) and (b), the reference case of $\chi=0$ is shown by the gray dashed curve for $\tilde a = 1$. 
}
\label{fig:pressure}
\end{figure}

Figure \ref{fig:pressure}\subref{fig:pressure_a} shows the effective interaction pressure acting on randomly charged surfaces ($\tilde \xi=0$) with mobile rodlike counterions in between as obtained from Eqs. \eqref{eq:f_plates} and \eqref{eq:p_plates}. The results are shown in rescaled units as a function of the intersurface distance for different values of the disorder coupling parameter, $\chi$, as indicated on the graph, at fixed $\tilde t=1$ and $\tilde a=1$ (colored solid curves). We also show the interaction pressure for the same parameter values by setting $\tilde t=0$ (colored dashed curves). For comparison, the reference case of counterions between uniformly charged surfaces ($\chi=0$) is also plotted as a gray dashed curve, which is analytically obtained as 
\begin{equation}
\tilde P_0(\tilde d)=-1+\frac{2}{\tilde d-2\tilde a},  
\label{eq:P0}
\end{equation}
which, as noted before, coincides with the result in the case of a monopolar system of counterions ($\chi=\tilde t=0$), provided other system parameters are kept the same. $\tilde P_0(\tilde d)$ decreases monotonically with $\tilde d$ and becomes attractive for $\tilde d>2+2\tilde a$. This attraction is a direct consequence of the strong-coupling electrostatics due to multivalent counterions as extensively reviewed before \cite{book,holm,hoda_review,Naji_PhysicaA, Naji2010, perspective, asim}.

\floatsetup[figure]{style=plain,subcapbesideposition=top}

The differences between  the colored dashed curves ($\tilde t=0$) and the reference, gray dashed, curve thus represent the effects, stemming merely from the surface charge disorder, without accounting for the additional quadrupolar effects, that can be identified by the differences found between the colored dashed curves and the corresponding solid curves (same $\chi$ values) in the figure.  One can also infer that, while the disorder alone can produce a more attractive (negative) pressure profile, with a mildly nonmonotonic shape, the combination of quadrupolar and disorder effects produces a much more attractive profile across the range of small to intermediate intersurface separations, with a pronounced local minimum, whose location shifts toward small values of intersurface separation, as $\chi$ is increased. The pressure profiles for $\tilde t>0$ (solid) and $\tilde t=0$ (dashed curves) merge at sufficiently large separations, indicating that the large-separation behavior of the interaction pressure is primarily determined by the surface charge distribution and the net monopolar charge of the counterions, which are kept the same for the different curves shown in the plot. 

The maximum intersurface attraction pressure can be larger, by more than an order of magnitude in the figure, than the largest value obtained in the reference, nondisordered, case as $\tilde P_0\rightarrow -1$, when $\tilde d\rightarrow\infty$ (this limiting pressure can be as large as a few atm in actual units; see Appendix \ref{app:parameters}). 

In the situation where surfaces bear disordered charge distributions of finite in-plane correlation length, $\tilde \xi$, the attractive pressure mediated between the bounding surfaces decreases in magnitude as the correlation length is increased. Hence, the electrostatic effects due to charge disorder are generally weakened, approaching the reference case of no charge disorder as shown in Fig. \ref{fig:pressure}\subref{fig:pressure_b}. 

\section{Summary}
\label{sec:conclusion}

We have studied the strong-coupling electrostatics of mobile rodlike counterions (charged nanorods) confined within a slit region  between heterogeneously charged plane-parallel surfaces, immersed in a bathing polar solution. The counterions possess both monopolar and quadrupolar moments in their internal charge distribution. While the monopolar moment of counterions couples to both the mean charge and the heterogeneous (disordered) component of the bounding surface charge distributions, their quadrupolar moment is sensitive only to field variations produced in space by the disordered surface component. These coupling mechanisms can be characterized by the mean electrostatic coupling parameter and the disorder coupling parameter that scale with the mean surface charge density and the disorder variance, respectively; they also grow with the third and the second power of the monopolar charge valency of counterions. Hence, since charged nanorods are usually multivalent, their presence in the solution leads to strong electrostatic couplings that go beyond the domain of validity of the traditional mean-field (Poisson-Boltzmann) theories. The quadrupolar effects due to rodlike multivalent counterions were analyzed previously by some of the present authors in the absence of charge disorder \cite{JPCM2009}. 

The current work focuses on the role of surface charge disorder and places the electrostatics of charged nanorods within the context of the newly growing field of disorder-induced phenomena 
\cite{Meyer,Meyer2,klein,klein1,klein2,kekicheff1,kekicheff2,kekicheff3,kantor-disorder0,andelman-disorder,Bing1,Bing2,Rabin,netz-disorder2,Lukatsky1,Lukatsky2,ali-rudi, rudiali, partial, Ghodrat1, Ghodrat2, Ghodrat3, Andelman2013, Andelman2016, Andelman2017, Dietrich2018,Hribar,Shklovski-07,disorder-PRL,jcp2010,pre2011,epje2012,jcp2012,speake,kim2,kim3}, especially in their strong-coupling manifestation in the soft-/biomatter context, where counterintuitive effects such as antifragility have recently been reported \cite{Ghodrat1,Ghodrat2,Ghodrat3}. 

We have studied the details of the interplay between the strong-coupling electrostatics, quadrupole charge distribution effects and disorder-induced counterion-surface interactions and have shown how this interplay gives rise to an increased accumulation of rodlike counterions near randomly charged surfaces, where the quadrupolar counterions are found to exhibit an orientational nematic order in the direction normal to the surface planes, an effect which is maximized when the surface disorder has no in-plane correlations. Orientational ordering of this type is in stark contrast with the typical parallel-to-surface orientation found for rodlike counterions at uniformly charged boundaries \cite{perspective, JPCM2009}, and is quite relevant for the understanding and control  of the deposition of charged rodlike viruses and viruslike nanoparticles \cite{assembly1, assembly2, assembly3, assembly4} on charged planar substrates in the formation of ordered films \cite{assembly4,assembly3, assembly1}. Apart from the importance of electrostatic interactions, the role of the disorder in these systems is crucial as the deposition preparation method unavoidably introduces frozen charge disorder effects into the final assembly of the materials. Our analysis clearly shows that the combined effects of electrostatics and charge disorder introduces qualitative changes in the behavior of the adsorbed phase, modifying the nature of the orientational order. Focusing on the characterization and modulation of the surface charge disorder, one can easily envision an orientational ordering transition, between the in-plane to perpendicular orientational ordering of the adsorbate, that would also modify  the interactions between two closely apposed adsorbing surfaces. 

We have shown that the effective interaction pressure between randomly charged surfaces shows an intriguing nonmonotonic behavior with a deep local minimum, representing a strongly negative interaction pressure at sufficiently large disorder strengths. The magnitude of this attractive pressure can be several times (or, depending on the parameter values, even more than an order of magnitude) larger than the usual strong-coupling attractive pressure, engendered by multivalent counterions between uniformly charged surfaces  \cite{book,holm,hoda_review, Naji_PhysicaA, Naji2010, perspective}, which again qualitatively changes the properties of these intersurface interactions. Such strong variations in the interactions between charge disordered surfaces, as first uncovered within a different context \cite{Ghodrat1,Ghodrat2,Ghodrat3}, point to the conclusion that the disorder, which is a constitutive feature of many systems involving bathing-solution-exposed bounding surfaces, should be viewed more as an opportunity to engineer their behavior than something one should try to avoid at all costs.

Our results are based on a few simplifying assumptions that can be improved within the same theoretical framework as used in this work. These include the image charges from the interfacial dielectric discontinuities, the added salt, and the finite spatial extension of the counterions  \cite{Forsman06,Messina,bohinc1,bohinc-rev,bohinc2012,Woon1,Woon2,Woon3,Woon4,JPCM2009,perspective}. These factors, especially the dielectric/salt  image charges, are expected to play significant roles in the interfacial regions close to the boundaries \cite{SCdressed2,SCdressed3,JPCM2009,rudiali,Ghodrat1,Ghodrat2,Ghodrat3},  thus making our current results to be valid strictly in the limit of vanishing dielectric discontinuity, and/or in situations where the charged surfaces have a small thickness \cite{jho-prl}. The presence of an added monovalent salt in the system can be handled using the dressed multivalent-ion approach  \cite{Ghodrat1,Ghodrat2, Ghodrat3, perspective,SCdressed1,SCdressed2,SCdressed3}, in which  multivalent ions are treated using strong-coupling schemes, while monovalent salt ions are treated using weak-coupling schemes, which will be discussed in the case of charged nanorods elsewhere \cite{to-be-published1}. Finally, the surface charge can show variable degrees of quenched and annealed behavior \cite{partial,Hribar}, constituting another interesting direction for future investigation of charged nanorods at heterogeneously charged boundaries.  

\begin{acknowledgments}
A.N. acknowledges partial support from Iran Science Elites Federation and the Associateship Scheme of The Abdus Salam International Centre for Theoretical Physics (Trieste, Italy), and thanks the University of Chinese Academy of Sciences and the Institute of Physics, Chinese Academy of Sciences, for their support and hospitality during a scientific visit leading to the completion of the final form of this paper. K.H. and E.M. thank the School of Physics, IPM, for hospitality and support during the completion of this work. R.P. was supported by the 1000-Talents Program of the Chinese Foreign Experts Bureau.
\end{acknowledgments}

\vspace{-3mm}

\appendix

\section{Overview of strong-coupling approach}
\label{app:SC}

For a given realization $\rho = \rho(\mathbf{r})$ of the quenched charge distribution, the Hamiltonian of the system under consideration can be written as 
\begin{equation}
\mathcal{H}_N= \frac{1}{2} \int \!\rmd\mathbf{r} \rmd\mathbf{r}' \big[ \hat{\rho}_N(\mathbf{r}) + \rho(\mathbf{r}) \big] G_0(\mathbf{r},\mathbf{r}') \big[ \hat{\rho}_N(\mathbf{r}') +\rho(\mathbf{r}') \big], 
\label{eq:hamiltonian_tot}
\end{equation}
with the implicit assumption that the single-particle self-energy terms are excluded. These terms are of no consequence in the present context, where there are no dielectric or salt inhomogeneities in the system (see, however, Refs. \cite{SCdressed2,SCdressed3,JPCM2009,rudiali,Ghodrat1,Ghodrat2,Ghodrat3}). They represent formation energy of individual counterions in free space and  are thus independent of particle position and orientation; they can safely be dropped or absorbed into the fugacity. We have also defined the total charge density operator of counterions as  $\hat{\rho}_N(\mathbf{r})=\sum_{i=1}^N \hat{\rho}_{c}(\mathbf{r}; \mathbf{R}_i, \boldsymbol{\omega}_i)$, where $\hat{\rho}_{c}(\mathbf{r}'; \mathbf{r}, \boldsymbol{\omega})$ is the single-particle charge density operator,  Eq. \eqref{eq:one-particle-op}. 

 The canonical partition function of the system with $N$ rodlike counterions thus reads  
\begin{equation}
{\mathcal Z}_N[\rho] =  \frac{1}{N!}\int \left(  \prod_{i = 1}^{N}\frac{\rmd\mathbf{r}_i}{\lambda_t^3}\rmd\boldsymbol{\omega}_i\, \Omega(\mathbf{r}_i,\boldsymbol{\omega}_i)\! \right)\rme^{- \beta \mathcal{H}_{N}}, 
\end{equation}
where $\lambda_t$ is the thermal wavelength of counterions. An efficient field-theoretic representation can be obtained for the grand-canonical partition function, 
 ${\mathcal Z}_\lambda = \sum_{N = 0}^{\infty} \lambda_0^{N} {\mathcal Z}_N$ (with $\lambda_0$ being the bare fugacity), through a standard Hubbard-Stratonovich transformation over the fluctuating potential field $\varphi(\mathbf{r})$  \cite{Netz01,holm,book,Edwards,Podgornik89,Podgornik89b,Netz-orland,JPCM2009}. We  arrive  (up to an irrelevant prefactor) at
\begin{equation}
{\mathcal Z}_\lambda[\rho]= \int \frac{\mathcal{D}\varphi}{\sqrt{\det G}} \,\rme^{ -\beta S_\lambda[\varphi; \rho]}.
\end{equation}
The effective electrostatic field-action is given by
\begin{align}
&S_\lambda[\varphi;\rho] =  \frac{1}{2}\int\! \rmd\mathbf{r} \rmd\mathbf{r}' \, \varphi (\mathbf{r}) G_0^{-1}(\mathbf{r}, \mathbf{r}') \varphi (\mathbf{r}') + \icomplex\! \int\! \rmd\mathbf{r} \,\rho(\mathbf{r}) \varphi(\mathbf{r}) \nonumber\\
&- \lambda \kBT \int \rmd\mathbf{r}\rmd\boldsymbol{\omega}\, \Omega(\mathbf{r},\boldsymbol{\omega})\, \rme^{-\icomplex \beta \!\int \rmd\mathbf{r}'\, \hat{\rho}_{c}(\mathbf{r}'; \mathbf{r}, \boldsymbol{\omega}) \varphi(\mathbf{r}')}, 
\label{eq:action_app}
\end{align}
with $ G_0^{-1}(\mathbf{r}, \mathbf{r}') = -\varepsilon_0 \varepsilon_m  \nabla^2 \delta(\mathbf{r} - \mathbf{r}')$ being the inverse Coulomb kernel and $\lambda=\lambda_0/\lambda_t^3$  the redefined fugacity.  

In the presence of quenched disorder \cite{dotsenko1,dotsenko2}, the thermodynamic free energy of the system follows by averaging the sample free energy (as opposed to averaging the sample partition function itself as would be required for annealed disorder \cite{partial,disorder-PRL,jcp2010,jcp2012}) over the disorder field as  ${\mathcal F}_\lambda = - \kBT\,\langle\!\langle\ln {\mathcal Z}_\lambda[\rho] \rangle\!\rangle$. After the necessary calculations (e.g., explicit evaluation of the partition function, ${\mathcal Z}_\lambda[\rho]$) are done, the results can be transformed back to the canonical ensemble  by fixing the fugacity $\lambda=\lambda(N)$ from the standard relation $N = -\beta \partial{\mathcal F}_\lambda/\partial \ln\lambda$, where $N$ is set by  the electroneutrality condition \eqref{eq:ENC}. Hence, the properly averaged free energy of the canonical system is obtained by Legendre transformation as ${\mathcal F}_N = {\mathcal F}_{\lambda(N)} + N \kBT\ln\lambda(N)$. 

The calculation of ${\mathcal Z}_\lambda[\rho]$ seldom yields itself to exact solution \cite{Edwards,exact1,exact2} rather than approximate methods or numerical simulations. As mentioned in the text, since charged nanorods typically possess a large net charge valency, one can use a virial expansion, in combination with a $1/\Xi$ expansion, to derive the limiting, strong-coupling, behavior of the system \cite{Netz01}. Despite its asymptotic nature, the strong-coupling theory and its more recent generalizations \cite{perspective} provide a simple, and yet powerful, analytical framework applicable over a wide range of realistic parameter values, as shown by numerous simulations  \cite{book,holm,AndrePRL,AndreEPJE,hoda_review,Naji_PhysicaA,Naji2010,perspective,asim,SCdressed1,SCdressed2,SCdressed3,NajiNetzEPJE,NajiArnold,Naji_CCT1,Naji_CCT2,Weeks,Weeks2,Jho1,arnoldholm,jho-prl,asim,Woon1,Woon2,Woon3,Woon4}  and also experiments \cite{hoda_review,Trefalt}. The limiting theory can nevertheless be improved using systematic calculations of the higher-order corrections \cite{Netz01,AndrePRL,AndreEPJE,trizac}, or through alternative approximation methods \cite{Santangelo,Hatlo-Lue,Forsman04,Burak04}. 

To establish the limiting theory in the present context, we virial-expand the grand-canonical partition function as $ {\mathcal Z}_\lambda[\rho]\simeq  {\mathcal Z}_0[\rho]+\lambda {\mathcal Z}_1[\rho]+{\mathcal O}(\lambda^2)$, giving the leading-order sample free energy 
$\beta {\mathcal F}_\lambda[\rho] = - \ln {\mathcal Z}_\lambda[\rho]= -\ln {\mathcal Z}_0[\rho]-\lambda {{\mathcal Z}_1[\rho]}/{{\mathcal Z}_0[\rho]}$, with
\begin{align}
{\mathcal Z}_0[\rho]&= \exp\left[-\frac{\beta}{2} \! \int \!\rmd\mathbf{r} \rmd\mathbf{r}' \, \rho (\mathbf{r}) G_0(\mathbf{r}, \mathbf{r}') \rho (\mathbf{r}') \right],
\label{eq:Z_0}
\\
{\mathcal Z}_1[\rho]&=  {\mathcal Z}_0[\rho] \int \rmd{\mathbf R} \rmd\boldsymbol{\omega}\, \Omega({\mathbf R},\boldsymbol{\omega})
\label{eq:Z_1}
\\
&\quad\times \exp\left[ - \beta\! \int\!\rmd\mathbf{r} \rmd\mathbf{r}' \hat{\rho}_{c}(\mathbf{r}; {\mathbf R}, \boldsymbol{\omega})  G_0(\mathbf{r}, \mathbf{r}')  \rho(\mathbf{r}')  \right]. 
\nonumber
\end{align}
(Note that the previously mentioned particle formation energies, if absorbed within the fugacity, will now be cancelled out by similar factors arising through the virial expansion; i.e., $\lambda$ in the above expressions will coincide with the redefined fugacity introduced after Eq. \eqref{eq:action_app} \cite{JPCM2009,ali-rudi}.) The above expressions can be averaged over the (Gaussian) disorder field using Eqs. \eqref{eq:rho-mean} and \eqref{eq:rho-variance}. One then obtains the thermodynamic free energy ${\mathcal F}_\lambda=\langle\!\langle{\mathcal F}_\lambda[\rho] \rangle\!\rangle$ as 
\begin{align}
 {\mathcal F}_\lambda&=  \frac{1}{2} \int \rmd\mathbf{r} \rmd\mathbf{r}'\,\rho_0(\mathbf{r})G_0(\mathbf{r},\mathbf{r}')\rho_0(\mathbf{r}')
\nonumber\\
&+ \frac{1}{2} \int\rmd\mathbf{r} \rmd\mathbf{r}'\, {\mathcal G}(\mathbf{r},\mathbf{r}')G_0(\mathbf{r},\mathbf{r}')
\nonumber\\
&- \lambda \kBT\int \rmd{\mathbf R} \rmd\boldsymbol{\omega}\, \Omega({\mathbf R},\boldsymbol{\omega})\,  \rme^{ -\beta u\left({\mathbf R}, \boldsymbol{\omega} \right)},  
\label{eq:F_lambda}
\end{align}
with $u({\mathbf R}, \boldsymbol{\omega})$ obtained as in Eqs. \eqref{eq:u_general}-\eqref{eq:u_dis}. Upon the transformation back to the canonical ensemble, we find the fugacity
\begin{equation}
\lambda = \frac{N}{\int \rmd{\mathbf R} \rmd\boldsymbol{\omega}\, \Omega({\mathbf R},\boldsymbol{\omega})\,  \rme^{ -\beta u\left({\mathbf R}, \boldsymbol{\omega} \right)}}, 
\label{eq:lambda}
\end{equation}
enabling one to reproduce the canonical expressions for the free energy, counterion density and  orientational order parameter as in Eqs. \eqref{eq:F_N}, \eqref{eq:c_general} and \eqref{eq:S_general}. 

\section{Effective dumbbell model}
\label{app:dumbbell}

While actual rodlike counterions have a finite length of $2\ell_0$, our model incorporates the rod length only through its monopolar and quadrupolar moments by introducing an effective rod length of $2\ell$, with  $\ell$  defined in Eq. \eqref{eq:ell_def_1}. This latter quantity can be interpreted as the half-length of an equivalent dumbbell, having the same monopolar and quadrupolar moments, $q$ and $t$, respectively. Such a charged dumbbell can be constructed by two pointlike end-caps each carrying a charge of $qe_0/2$ and being placed uniaxially and at equal distances $\ell$ on the two side of its center of charge $\mathbf{R}$, giving the  charge density operator 
 $\hat{\rho}_{c}(\mathbf{r}; \mathbf{R}, \boldsymbol{\omega}) = ({qe_0}/{2}) \big[\delta\big(\mathbf{r} - ( \mathbf{R} + \boldsymbol{\ell} ) \big) + \delta\big(\mathbf{r} - (\mathbf{R} - \boldsymbol{\ell} ) \big) \big]$. 
Expanding this expression in powers of $\boldsymbol{\ell} = \ell \mathbf{n}$ gives 
\begin{equation}
\hat{\rho}_{c}(\mathbf{r}; \mathbf{R}, \boldsymbol{\omega}) = e_0 \sum_{j=0}^\infty t^{(2j)} \big( \mathbf{n} \cdot \nabla \big)^{2j} \delta\left(\mathbf{r} - \mathbf{R} \right), 
\end{equation}
where the nonzero multipole moments $t^{(2j)}$ are defined in accordance with the general expression \eqref{eq:one-particle-op} as 
\begin{equation}
t^{(2j)} \equiv q\frac{\ell^{2j}}{(2j)!}  \quad\quad j=0,1,2,\cdots. 
\end{equation}
The quadrupolar moment is then expressed as
\begin{equation}
t \equiv t^{(2)} = q\frac{\ell^2}{2}, 
\end{equation}
reproducing Eq. \eqref{eq:ell_def_1} in the text. 

\section{Choice of parameter values}
\label{app:parameters}

Since our results are given in rescaled form, they can be mapped to a wide range of actual parameter values, provided the validity criteria for the current theoretical framework are fulfilled  (Section \ref{subsec:validity_regime}). Here, we discuss the admissible ranges of values that can be chosen for the system parameters in actual units, provide numerical estimates for them based on a few real-life examples of charged nanorods, and show that those estimates will be consistent with the ranges of dimensionless parameter values used in Section \ref{sec:plane-parallel_num}. 

The strong-coupling criterion, Eq. \eqref{eq:sc}, gives a general necessary condition on any potential comparisons that may be attempted between the present results and experiments and/or computer simulations. This criterion sets an upper limit on the rescaled intersurface distance, $\tilde d=d/\mu$, or, conversely, a lower limit on $\Xi$. Let us assume that $\tilde d$ is varied over the interval $2\tilde a\leq \tilde d\leq {\tilde d}_{\mathrm{max}}$, where the current  theory is assumed to remain valid. Hence, according to Eq. \eqref{eq:sc}, we need to take 
\begin{equation}
\Xi > \tilde d^{\,2}_{\mathrm{max}}/4. 
\label{eq:constraint_d_max}
\end{equation}
We can conveniently express ${\tilde d}_{\mathrm{max}}$ as ${\tilde d}_{\mathrm{max}}=\varsigma \tilde \ell$. Because of the multipole-expansion criterion, Eqs. \eqref{eq:ell_less_a} and \eqref{eq:cutoff}, we need to take $\varsigma>\sqrt{2}$. However, the excluded-volume criterion, Eq. \eqref{eq:steric}, can be satisfied only if the range of values for $\varsigma$ are  further restricted as  $\varsigma>2$. This constraint is evidently fulfilled by our choice of $\tilde d_{\mathrm{max}} = 7\tilde a = 7\tilde \ell/\sqrt{2}$ or $\varsigma = 7/\sqrt{2}$ in Section \ref{sec:plane-parallel_num}. 

Using the definitions of the Gouy-Chapman length and the coupling parameter, Eqs. \eqref{eq:mu} and (\ref{eq:Xi}), the constraint \eqref{eq:constraint_d_max} can be written as $\sigma<2q/(\pi d^{\,2}_{\mathrm{max}})$   in actual units. Thus, for a given type of rodlike counterions with monopolar charge valency $q$ and effective half-length $\ell$ or, equivalently, the {\em effective linear charge density},  $\tau=q/(2\ell)$, the strong-coupling criterion is translated into a constraint on the surface charge density as
\begin{equation}
\sigma < \frac{4\tau}{\pi\varsigma^2\ell}. 
\label{eq:constraint_sigma}
\end{equation}
We emphasize again that the effective length, $2\ell$, and linear charge density, $\tau$, of rodlike counterions that appear in our model may in general be different from their actual (experimental) values, $2\ell_0$ and $\tau_0$, respectively. To compare with any particular type of rodlike counterions, the monopolar charge valency and the quadrupolar moment are to be adopted as the only model input parameters.

Short DNA fragments behave as stiff negatively charged nanorods (with $\tau_0\simeq 5.9$~${\mathrm{nm}}^{-1}$) when prepared in sizes smaller than their persistence length (typically 30-50~nm, depending on the ionic strength). Thus, for DNA nanorods of actual length $2\ell_0=10$~nm, and by adopting a uniformly-charged-rod model with $q= 2\tau_0\ell_0$ and $t=\tau_0\ell_0^3/3$, we find the effective half-length $\ell=\ell_0/\sqrt{3}\simeq 2.9$~nm and the linear charge density $\tau=\sqrt{3}\tau_0\simeq 10.2$~${\mathrm{nm}}^{-1}$. Using Eq. (\ref{eq:constraint_sigma}) and the previously mentioned value of $\varsigma$, we obtain the admissible values of $\sigma$ as $\sigma< 0.18$~${\mathrm{nm}}^{-2}$, which falls well within the experimentally accessible regime. 

As other potential examples, we consider (cationic) polyamines such as trivalent spermidine and tetravalent spermine. However, we should first note that the actual charge of these molecules varies depending on the solution $p$H (we assume complete protonation of the amine groups) \cite{Gimsa} and that their linear structure is generally flexible \cite{Bohinc2011rev,Feuerstein1986,Feuerstein1990}, even though they have also been modeled as rodlike polycations in the literature \cite{bohinc-rev,bohinc2012,Gimsa}, which will be cautiously adopted here as well. Modeling   spermidine (spermine) as a linear array of three (four) equi-distanced monovalent groups and actual end-to-end length of $2\ell_0 \simeq  1.09$~nm ($2\ell_0 \simeq  1.55$~nm) \cite{Bohinc2011rev,Gimsa,Feuerstein1986,Feuerstein1990}, we find the quadrupolar moment of  $t=\ell_0^2$~${\mathrm{nm}}^2$ ($t=10\ell_0^2/9$~${\mathrm{nm}}^2$) and the  effective half-length of $\ell \simeq 0.45$~nm ($\ell \simeq  0.58$~nm). Equation \eqref{eq:constraint_sigma} then gives $\sigma< 0.39$~${\mathrm{nm}}^{-2}$ ($\sigma < 0.31$~${\mathrm{nm}}^{-2}$), which is again well within the experimentally admissible range. 

In the foregoing examples, the upper limits on $\sigma$ can be converted to lower limits on the Gouy-Chapman length, $\mu=1/(2\pi q\ellB\sigma)$, or upper limits on $\tilde \ell = \ell/\mu$ as $\tilde \ell< 137$ (DNA nanorods), 2.3 (spermidine), and 3.2 (spermine), in an aqueous solvent at room temperature ($T=293$~K, $\varepsilon_m\simeq 80$, and $\ellB\simeq 0.71$~nm), consistently covering the range of values $\tilde \ell=\sqrt{2\tilde t}\leq 2$ used  in Section \ref{sec:plane-parallel_num}. 

It is also clear that, within the range of values chosen for the disorder coupling parameter $\chi$ (up to around $\chi=10$ in Section \ref{sec:plane-parallel_num}), the Gaussian criterion $g<\sigma$ (Eq. \eqref{eq:gauss}) can easily be satisfied for large valency counterions, as $g=\chi/(2\pi q^2\ellB^2)$ will have a relatively small value. For instance, for the cases of DNA, spermidine and spermine as discussed above, we find $g\simeq 9\times 10^{-4}$, 0.35, and 0.2~${\mathrm{nm}}^{-2}$ for fixed $\chi=10$, respectively. Smaller values of $g$ can be obtained by taking solvents, or solvent mixtures, with lower dielectric constants \cite{Hoagland03}.
For instance, using $\varepsilon_m=40$, we find a Bjerrum length twice that in water; hence, the estimated values of $g$ will be four times smaller, better fitting the Gaussian criterion. 

Finally, we note that for the typical range of $\sigma=0.1-0.2$~${\mathrm{nm}}^{-2}$, the characteristic pressure (see Eq. \eqref{eq:rescaled_defs}) is obtained as $2\pi \ellB \sigma^2/\beta\simeq 1.8-7.2$~atm. This 
can be used to convert the dimensionless pressure values reported in Section \ref{subsec:pressure}. 

\vspace{-5mm}


\end{document}